\documentclass[twocolumn,tighten,mathlines,numberedappendix,twocolappendix]{aastex631}

\usepackage{amsmath}

\def\simlt{\lower.5ex\hbox{$\; \buildrel < \over \sim \;$}}
\def\simgt{\lower.5ex\hbox{$\; \buildrel > \over \sim \;$}}

\def\beq{\begin{equation}}
\def\eeq{\end{equation}}
\def\bB{\boldsymbol{B}}
\def\bE{\boldsymbol{E}}

\def\bj{\boldsymbol{j}}

\def\bk{\boldsymbol{k}}

\def\Sect{{\rm Section}}

\def\Eq{Equation}
\def\Eqs{Equations}

\def\sT{\sigma_{\rm T}}

\def\Lw{L_{\rm w}}
\def\E{{\cal E}}
\def\N{{\cal N}}

\def\RLC{R_{\rm LC}}

\def\tobs{t_{\rm obs}}

\def\EFRB{{\cal E}_{\rm FRB}}

\def\tF{\tilde{F}}

\def\tn{\tilde{n}}

\def\bn{\boldsymbol{n}}

\def\bsh{\beta_{\rm sh}}

\def\tn{\tilde{n}}

\def\omB{\omega_B}

\def\gsh{\gamma_{\rm sh}}

\def\Bbg{B_{\rm bg}}
\def\bBbg{\boldsymbol{B}_{\rm bg}}
\def\Ebg{E_{\rm bg}}

\def\tBbg{\tilde{B}_{\rm bg}}
\def\tomB{\tilde{\omega}_B}
\def\tB{\tilde{B}}
\def\tE{\tilde{E}}

\def\gD{\gamma_{\rm D}}
\def\bD{\beta_{\rm D}}

\def\bs{\beta_{\rm s}}
\def\gs{\gamma_{\rm s}}

\def\trho{\tilde{\rho}}

\def\xiin{\xi_{\rm i}}

\def\c{\kappa}
\def\cu{\c_{\rm u}}
\def\cd{\c_{\rm d}}

\def\sigbg{\sigma_{\rm bg}}
\def\rhobg{\rho_{\rm bg}}
\def\nbg{n_{\rm bg}}

\def\xish{\xi_{\rm sh}}

\def\Em{E_0}
\def\Rx{R_{\times}}
\def\rsh{r_{\times}}

\def\sigx{\sigma_{\times}}

\def\gu{\gamma_{\rm u}}
\def\gd{\gamma_{\rm d}}

\def\betad{\beta_{\rm d}}

\def\tcross{t_{\rm cross}}

\def\sigu{\sigma_{\rm u}}

\def\eps{\epsilon}
\def\Bd{B_{\rm d}}
\def\Ed{E_{\rm d}}

\def\me{m}

\def\KF{{\cal K}}
\def\tKF{\tilde{\cal K}}

\def\Ep{E_{\rm p}}

\def\sigsc{\sigma_{\rm sc}}
\def\tausc{\tau_{\rm sc}}

\def\tBu{\tilde{B}}

 \def\Enpre{\EFRB}
\def\ompre{\omega_{\rm pre}}
\def\tompre{\tilde{\omega}_{\rm pre}}
\def\ompred{\mathring{\omega}_{\rm pre}}
\def\Epre{\delta E}

\def\tEpre{\delta\tilde{E}}

\def\Upre{U}
\def\tUpre{\tilde U}
\def\KFd{\mathring{{\cal K}}}

\def\Upred{\mathring{U}}
\def\Epred{\delta \mathring{E}}

\def\Bud{\mathring{B}}
\def\bEud{\mathring{\boldsymbol{E}}}
\def\bBud{\mathring{\boldsymbol{B}}}
\def\gud{\mathring{\gamma}}

\def\Bdd{\mathring{B}_{\rm d}}
\def\omd{\mathring{\omega}_{\rm d}}

\def\gshd{\mathring{\gamma}_{\rm sh}}

\def\Lpre{L_{\rm pre}}

\def\nupre{\nu_{\rm pre}}

\def\tu{\tilde{u}}
\def\tg{\tilde{\gamma}}
\def\dEe{\dot{\E}_e}
\def\xiB{\varepsilon}

\def\Wp{W_{\rm p}}

\def\gbg{\gamma_{\rm bg}}

\def\xrad{x_{\rm rad}}
\def\Rrad{R_{\rm rad}}

\def\gD{\gamma_{\rm D}}

\def\csh{\kappa_{\rm sh}}
\def\cshd{\mathring{\kappa}_{\rm sh}}
\def\cud{\mathring{\kappa}}
\def\bDud{\mathring{\beta}_{\rm D}}
\def\gDud{\mathring{\gamma}_{\rm D}}

\def\bcr{b_{\rm s}}

\def\C{{\cal C}}

\def\rem{r_{\rm em}}
\def\aem{a_{\rm em}}

\def\cbg{\kappa_{\rm bg}}
\def\gbg{\gamma_{\rm bg}}
\def\bbg{\beta_{\rm bg}}
\def\sigLC{\sigma_{\rm LC}}
\def\Erms{E_{\rm rms}}
\def\tErms{\tE_{\rm rms}}
\def\dErms{\mathring{E}_{\rm rms}}
\def\gem{\gamma_{\rm em}}
\def\Sigem{\Sigma_{\rm em}}

\def\Nem{\N_{\rm em}}
\def\cem{\kappa_{\rm em}}
\def\xirad{\xi_{\rm rad}}
\def\xia{\xi_{\rm a}}
\def\xib{\xi_{\rm b}}

\def\Rabs{R_{\rm abs}}
\def\gstoch{\tilde\gamma_{\rm stoch}}
\def\grad{\tilde\gamma_{\rm rad}}

\def\NLC{\N_{\rm LC}}

\def\SS{Q}

\defcitealias{Beloborodov23}{B23}
\defcitealias{Beloborodov26}{B26}

\newbox\grsign \setbox\grsign=\hbox{$>$} \newdimen\grdimen \grdimen=\ht\grsign
\newbox\simlessbox \newbox\simgreatbox \newbox\simpropbox
\setbox\simgreatbox=\hbox{\raise.5ex\hbox{$>$}\llap
     {\lower.5ex\hbox{$\sim$}}}\ht1=\grdimen\dp1=0pt
\setbox\simlessbox=\hbox{\raise.5ex\hbox{$<$}\llap
     {\lower.5ex\hbox{$\sim$}}}\ht2=\grdimen\dp2=0pt
\setbox\simpropbox=\hbox{\raise.5ex\hbox{$\propto$}\llap
     {\lower.5ex\hbox{$\sim$}}}\ht2=\grdimen\dp2=0pt
\def\simgt{\mathrel{\copy\simgreatbox}}
\def\simlt{\mathrel{\copy\simlessbox}}

\begin{document}

\title{Radio precursors of monster shocks: a mechanism for fast radio bursts from SGR~1935+2154}

% \correspondingauthor{Andrei M. Beloborodov}
 \email{amb@phys.columbia.edu}

\author[0000-0001-5660-3175]{Andrei M. Beloborodov}
\affiliation{Physics Department and Columbia Astrophysics Laboratory, Columbia University, 538  West 120th Street New York, NY 10027,USA}
\affil{Max Planck Institute for Astrophysics, Karl-Schwarzschild-Str. 1, D-85741, Garching, Germany}

\begin{abstract}
Kilohertz perturbations in active magnetars evolve into monster radiative shocks at radii $r\sim 10^8$\,cm. The shock generates X-rays and a semi-coherent radio precursor, which strongly interacts with the magnetospheric plasma ahead of the shock. We show that this interaction self-regulates the precursor emission and find its self-consistent frequency and luminosity. The precursor frequency falls in the GHz band and its production peaks when the shock expands to $r\approx 10^9$\,cm. The resulting GHz burst has a sub-millisecond duration and energy $\EFRB\approx 10^{34}\E_{38}^{0.2}$\,erg where $\E$ is the energy of the primary  magnetosonic disturbance that launched the shock. As the GHz burst propagates to the light cylinder $\RLC\sim 10^{10}$\,cm, it faces a threat of being absorbed by the magnetosphere. The burst escapes if the local plasma density at $\RLC$ is $\sim 30$ times lower than typically expected for active magnetars, so distant observers need some luck to see the radio burst. The shock X-rays follow the radio waves with a millisecond delay. Shocks from kilohertz disturbances with energies $\E\sim 10^{38}$\,erg generate X-ray and radio bursts similar to the activity detected in SGR~1935+2154. 
\end{abstract}

\vspace*{-5mm}


%#####################################################################

\section{Introduction}

Fast radio bursts (FRBs) are the most spectacular and puzzling phenomena in the radio sky \citep{Petroff19}. At least some of them are believed to come from magnetars, and weak FRBs have been detected from a known magnetar SGR~1935+2154 \citep{CHIME20,Bochenek20}.

The emission mechanism of FRBs from magnetars is not settled. Some clues are provided by two constraints:
\\
 (i) Only the compact region around the neutron star at radii $r\lesssim 10^8$\,cm (``inner magnetosphere'') has sufficient magnetic energy to power FRBs. This region is deep inside the light cylinder of a typical magnetar, $\RLC\sim 10^{10}\,$cm.
\\
(ii) Strong radio waves are damped in the outer dipole magnetosphere and the wind at radii up to $\sim 10^{11}$\,cm \citep{Beloborodov24,Beloborodov26}.
\\
These constraints require a magnetic explosion that relocates energy from $r\lesssim 10^8$\,cm to $r\gtrsim10^{11}$\,cm before the FRB is released into the surrounding plasma. If the process of FRB production begins at smaller radii, it should operate inside the explosion ejecta.

Recent work demonstrated how magnetic explosions occur in the perturbed magnetospheres of neutron stars. Active magnetars experience macroscopic magnetohydrodynamic (MHD) perturbations of two types, Alfv\'enic and magnetosonic, with frequencies in the kHz band. Explosions launched by  Alfv\'enic (shear) perturbations were demonstrated numerically by \cite{Parfrey13,Mahlmann23,Yuan20}. Magnetosonic (compressive) perturbations also evolve into explosions and launch monster radiative shocks with plasma Lorentz factors exceeding $10^5$ (\citealt{Beloborodov23}, hereafter \citetalias{Beloborodov23}). The shock forms where the MHD wave approaches the condition $E^2=B^2$, which occurs at radius $\Rx\sim 10^8L_{41}^{-1/4}$\,cm 
 (\citealt{Chen22b}; \citetalias{Beloborodov23}; \citealt{Grehan26}) where $L$ is the power of the kHz disturbance. The shock converts $\sim$ half of the kHz perturbation energy into X-rays.

The magnetospheric shocks involve a jump of the magnetic field and an ultrarelativistic motion of the plasma. The jump structure of such magnetized shocks is unstable on the Larmor timescale, generating a semi-coherent, linearly polarized, radio precursor (e.g. \cite{Sironi21}). Previously, the precursors from shocks at large radii, far outside the magnetosphere, were proposed as FRB sources \citep{Lyubarsky14,Beloborodov17b,Beloborodov20}. Magnetospheric monster shocks are expected to emit similar precursors, possibly with a similar efficiency \citep{Vanthieghem25a,Bernardi25}. 

The present paper calculates the radio emission from monster shocks inside the magnetosphere. The calculation demonstrates a key feature: the radio precursor strongly interacts with the upstream magnetospheric plasma approaching the shock. This interaction dramatically reduces the shock strength and changes how the precursor itself is emitted. We investigate the self-regulated evolution of the ``shock + precursor'' structure, find the radius $\Rrad$ where the precursor emission peaks, and evaluate the self-consistent frequency and power of radio bursts produced by monster shocks.

The solution will show that the magnetospheric shocks radiate GHz bursts at $\Rrad\approx 10^9$\,cm with energies $\sim 10^{34}$\,erg. This result explains the weak FRBs from SGR~1935+2154. The proposed model differs from previous discussions of SGR~1935+2154, which assumed GHz emission from small radii $r<10^8$\,cm \citep{Lu20} or considered shock waves at large $r>10^{11}$\,cm \citep{Margalit20,Yuan20}.

The paper is organized as follows. Section~\ref{monster} describes monster shocks and their evolution using a simple equation, which accurately reproduces the results of full MHD simulations. Then, \Sect~\ref{precursor} focuses on the self-regulated emission of the radio precursor. Sample numerical solutions are shown in \Sect~\ref{numerical}, and an analytical description of the produced FRBs is given in \Sect~\ref{analytical}. Damping of the emitted FRB on its way out of the magnetosphere is discussed in \Sect~\ref{damping}. The results are summarized in \Sect~\ref{discussion}. Our main numerical example will be tailored to SGR~1935+2154, which has the magnetic dipole moment $\mu\approx 2\times 10^{32}$\,G\,cm$^3$ and rotation period $P=3.25$\,s.

%#################################################################

\section{Dynamics of the monster shock}
\label{monster}

\subsection{Setup of the problem}

At radii $r<\Rx$, the outgoing kHz magnetosonic perturbation behaves as a vacuum electromagnetic wave linearly superimposed on the background magnetosphere. The wave has a toroidal electric field $\bE\parallel (\bBbg\times\bk) $ where $\bk$ is the (radial) wavevector, and $\bBbg$ is the background magnetic field, approximated as dipole.

The kHz wave expands with the speed of light $c$ carrying an electric field profile $E(\xi)$, where 
\beq
  \xi\equiv t-\frac{r}{c}. 
\eeq
As an example, we will consider the wave profile 
\beq
  E(\xi)=E_0\sin(\omega \xi), \qquad 0<\xi<\frac{3\pi}{\omega}.
\eeq
Hereafter our convention is that $E>0$ corresponds to $B>\Bbg$ (compression phase of the wave) and $E<0$ corresponds to $B<\Bbg$ (rarefaction phase, see \citetalias{Beloborodov23}).
The average power of one oscillation ($0<\xi<2\pi/\omega$) is $L=c r^2 \Em^2/2$. The wave amplitude evolves as $E_0\propto r^{-1}$.  

The vacuum-like propagation of the electromagnetic perturbation ends where $B^2-E^2$ approaches zero. This first happens in the equatorial plane of the magnetosphere at radius $\Rx$ where $\Bbg=2E_0$:
\beq 
\label{eq:Rx}
  \Rx=\left(\frac{c\mu^2}{8L}\right)^{1/4}
   \approx 1.39\times 10^8\, \frac{\mu_{33}^{1/2}}{L_{43}^{1/4}}{\rm ~cm},
\eeq
where $\mu=r^3\Bbg$ is the magnetic dipole moment. The monster shock is launched at $\Rx$.
As shown in \citetalias{Beloborodov23}, the shock strength is controlled by the magnetization parameter of the unperturbed background magnetosphere,
\beq
\label{eq:sigbg}
    \sigbg\equiv \frac{\Bbg^2}{4\pi \rhobg c^2} =  \frac{\mu^2}{4\pi \me c^2 \N r^3},
    \qquad \N\equiv r^3\nbg,
\eeq
where $\rhobg=\me \nbg$ is the background plasma density, and $\me$ is the electron/positron mass. Simple estimates suggest that the parameter $\N$ is roughly constant with $r$, and its typical value is $\N\sim 10^{37}$ \citep{Beloborodov20}. The magnetization parameter at $\Rx$ is 
\beq
   \sigx\equiv\sigbg(\Rx)
   \approx 3.6\times 10^{9}\,\frac{\mu_{33}^{1/2}L_{43}^{3/4}}{\N_{37}}.
\eeq
The shock develops suddenly when the wave reaches $R_\times$. It forms in the rarefaction part of the magnetosonic perturbation where $B$ is minimum. The shock was named ``monster'' because the plasma passes through the shock with a huge Lorentz factor $\gamma\sim \sigx c/\omega\Rx$.

The present paper will focus on waves near the equatorial plane of the dipole magnetosphere, where $\bBbg$ is perpendicular to the wave propagation direction.\footnote{The results may be extended to oblique $\bBbg$ encountered by the radially expanding wave at different polar angles $\theta\neq \pi/2$. It forms a shock at a larger radius $\rsh(\theta)>\Rx$ \citepalias{Beloborodov23}.}
In the equatorial waves, $\bB$ remains parallel to $\bBbg$ (along the $\theta$-direction in spherical coordinates $r,\theta,\phi$), and the plasma executes the $\bE\times\bB$ drift with a radial velocity,
\beq
   \beta\equiv \beta_r=\frac{E}{B}, \qquad E\equiv -E_\phi, \quad B\equiv B_\theta.
\eeq 
The plasma drifts with Lorentz factor $\gamma=(1-\beta^2)^{-1/2}$.

The evolution of a kHz magnetosonic perturbation and shock formation can be calculated analytically if the perturbation frequency satisfies $\omega\gg c/\Rx$ \citepalias{Beloborodov23}. In such waves, the magnetic field is related to the electric field by
\beq
  B=\Bbg+E. 
\eeq
Furthermore, the continuity equation for plasma density $n$ and magnetic flux freezing give relations similar to those in plane-parallel waves:
\beq
\label{eq:continuity}
  \frac{B}{\Bbg}=\frac{n}{\nbg}=\frac{1}{1-\beta}, \quad 
  \sigma\equiv \frac{B^2}{4\pi n\gamma \me c^2} =\c\sigbg,
\eeq
where
\beq
\label{eq:kappa}
   \kappa\equiv\gamma(1+\beta)=\frac{1}{\gamma(1-\beta)}.
\eeq 
This parameter describes the plasma motion (note that $\c<1$ if $\beta<0$). It also equals the compression factor measured in the local fluid rest frame:
\beq
\label{eq:compr}
    \frac{\tn}{\nbg}=\frac{\tB}{\Bbg}=\c.
\eeq 

At a given time $t$, all parts of the short wave are located at nearly the same radius $r$.  Therefore, $r/c$ may be used as an approximate time coordinate:
\beq
   \frac{r}{c}\approx t \qquad \left(\xi\ll \frac{r}{c}\right).
\eeq
Similar to \citetalias{Beloborodov23}, we will use the spacetime coordinates $r,\xi$ instead of $t,r$: $r\approx ct$ is the radius of the short magnetosonic pulse at time $t$, and $\xi$ specifies the position within the pulse.

\subsection{Plasma acceleration in a kHz magnetosonic wave}
\label{plateau}

Evolution of a magnetosonic perturbation at $r\simgt \Rx$ may be summarized as follows. The wave profile $E(\xi)$ where $E^2<B^2$ continues to propagate nearly as in vacuum without significant deformation, and the ``offensive'' part of the profile where vacuum propagation would give $E^2>B^2$ is shaved off, keeping $E^2\approx B^2$ (see Figure~5 in \citetalias{Beloborodov23}). This creates a plateau of $E$:
\beq
\label{eq:Ep}
   E\approx \Ep=-\frac{\Bbg}{2} \qquad ({\rm plateau}),
\eeq
which corresponds to $E^2\approx B^2$. Note that $\Ep\propto r^{-3}$ while vacuum propagation would give $E\propto r^{-1}$. The difference from vacuum propagation is supported by
electric current $\bj$ along $\bE$,
\beq
\label{eq:jp}
       j=\frac{\partial_t(rE)_{\xi}}{2\pi r}\approx -\frac{c\Bbg}{2\pi r} \qquad ({\rm plateau}).
\eeq

Hereafter the beginning and end points of the plateau $E(\xi)\approx\Ep$ will be denoted as $\xiin$ and $\xish$ respectively (the monster shock is located at $\xish$).
The plateau occupies the region of width $\Wp$, which expands with time:
\beq
\label{eq:Wp}
  \xiin<\xi<\xish, \qquad \Wp=c(\xish-\xiin).
\eeq
The plateau expands at $r>\Rx$ because $\Ep/\Em\propto r^{-2}$, so $\Wp$ is forced to occupy an increasing fraction of the wave profile, quickly growing to $\Wp\sim c/\omega$.

\begin{figure}[t]
% \vspace*{-1.7mm}
\includegraphics[width=0.44\textwidth]{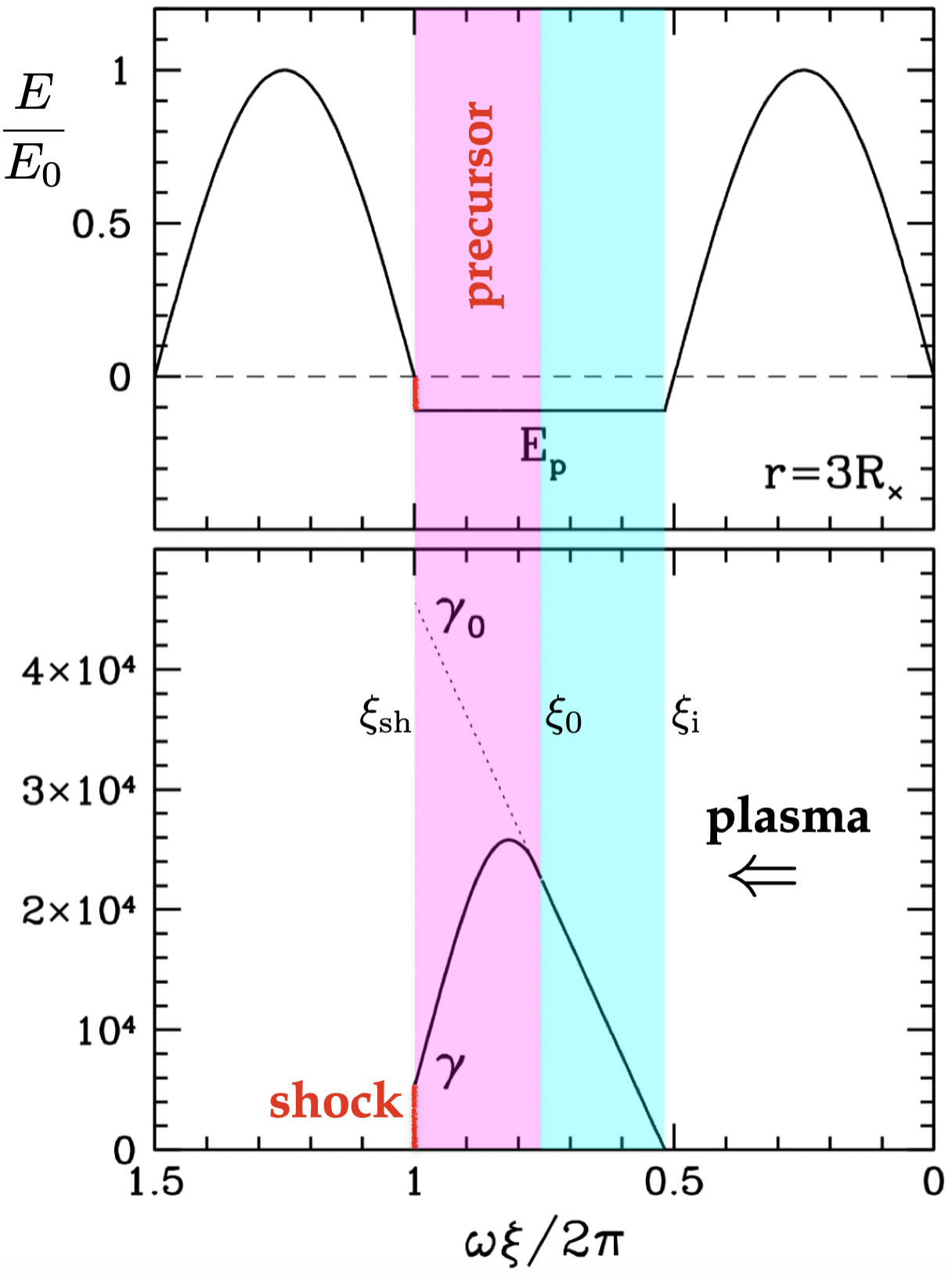} 
\caption{Snapshot of a kHz magnetosonic wave, taken when it has reached radius $r=3\Rx$. The wave profile is shown in the comoving coordinate $\xi=t-r/c$. The front of the wave is at $\xi=0$;  $\xi$ increases to the left. The monster shock emerged earlier, at $r=\Rx$, near the trough of the wave $\xi_0=3\pi/2\omega$. By the time of this snapshot, the shock position $\xish$ has reached $2\pi/\omega$, which corresponds to $\cd=1$. The upper panel shows the kHz wave electric field, which has evolved from its initial profile $E=E_0\sin(\omega\xi)$ by developing the plateau $E=\Ep\approx-\Bbg/2$, highlighted in cyan and purple. The purple region is occupied by the radio precursor --- the GHz FRB generated by the shock. The plasma particles are accelerated along the plateau to Lorentz factor $\gamma$ shown in the lower panel. The upstream flow would develop $\gamma_0$ (dotted line) if there were no precursor. The actual $\gamma$ (solid curve) is reduced by radiative losses induced by the precursor. The self-consistent calculation of $\gamma(\xi)$ and the precursor emission by the shock is explained in the next sections. The parameters of this sample model (Model~W) are given in \Sect~\ref{numerical}.}
\label{fig:plateau}
\end{figure}

As an example, consider a magnetospheric perturbation with initial $E(\xi)=\Em\sin(\omega\xi)$ (Figure~\ref{fig:plateau}). The offensive part of the initial wave profile is where $|E/\Ep|>1$, which occurs at $\sin(\omega\xi)<-\Rx^2/r^2$. It first appears at $r=\Rx$ near $\xi_0=3\pi/2\omega$ (i.e. in the trough of the wave where $B=\Bbg+E$ is minimum). Then, at $r>R_\times$, the plateau around $\xi_0=3\pi/2\omega$ widens. Its boundary $\xiin$ is located in the interval $\pi<\omega\xiin<3\pi/2$ and satisfies $\sin(\omega\xiin)\approx\Rx^2/r^2$, so 
\beq
\label{eq:xiin}
    \omega\xiin=\pi + \arcsin\left(\frac{\Rx^2}{r^2}\right).
\eeq

The plasma always moves in the $\xi$-coordinate with $d\xi/dt=1-\beta>0$ and crosses the plateau from $\xiin$ toward $\xish>\xiin$. The plasma drift in the $r$ coordinate has speed $\beta=E/B$, which approaches $-1$ on the plateau. Note that $\bE_{\rm p}\downarrow\uparrow (\bBbg\times\bk)$ while $\bB \uparrow\uparrow\bBbg$, so the plasma motion in coordinate $r$ (the $\bE\times \bB$ drift) occurs in the $-\bk$ direction (see Figure~2 in \citetalias{Beloborodov23}). Thus, the magnetosonic wave accelerates the plasma {\it toward} the star, $\beta<0$, like a  tsunami drawback. The acceleration ends with sudden deceleration in the shock at $\xish$. The shock is the jump from the plateau ($E^2\approx B^2$) to the undeformed part of the magnetosonic wave that still sustains vacuum-like propagation ($E^2<B^2$). 

The plateau accelerator and the shock are demonstrated in \citetalias{Beloborodov23} by the explicit MHD solution tracking the kHz magnetosonic wave propagation to $r>R_\times$. The solution can be described analytically. The plasma acceleration on the plateau obeys the energy equation,
\beq
\label{eq:energy}
   \me c^2 n\, \frac{d\gamma}{dt}=Ej= \frac{c\Bbg^2}{4\pi r}.
\eeq
Using $n=\nbg/(1-\beta)$ (\Eq~\ref{eq:continuity}) and $d\xi=(1-\beta)dt$ along the plasma streamline, one obtains from \Eq~(\ref{eq:energy}) the profile of $\gamma(\xi)$ along the plateau:\footnote{The analytical solution for $\gamma(\xi)$ holds in the region of $c(\xi-\xiin)\gg r/\sigbg$, i.e. practically on the entire plateau when $\Wp\gg r/\sigbg$. This condition corresponds to $\gamma\gg 1$ and becomes satisfied almost immediately after plateau formation at $\Rx$.}
\beq
\label{eq:gamma}
   \frac{d\gamma}{d\xi}=\frac{c\sigbg}{r} \quad \Rightarrow \quad 
   \gamma(\xi)\approx \frac{c\sigbg}{r}(\xi-\xiin).
\eeq 
As the plateau width grows to $\Wp\sim c/\omega$, the plasma flow in the kHz wave is pushed to a huge Lorentz factor,\footnote{We will show below that the precursor emitted by the shock occupies the second part of the plateau $\xi_0<\xi<\xish$ and hinders the flow acceleration at $\xi>\xi_0$ (Figure~\ref{fig:plateau}). Then, the linear accelerator acts unimpeded only in the first half of the plateau, $\xiin<\xi<\xi_0$, reaching the maximum Lorentz factor $\gamma(\xi_0)$.}
\beq
\label{eq:gu}
   \gamma\sim \frac{c\sigbg}{\omega r} =  \frac{\mu^2}{4\pi \me c \N r^4 \omega}
   \approx 4.6\times 10^3 \frac{\mu_{33}^2}{\N_{37}\nu_4 r_9^4}. 
\eeq
The relativistic flow at $\xi<\xish$ forms the upstream of the monster shock. The plateau accelerator emerges immediately at $r\approx\Rx$, reaches the peak $\gamma$ at $r\approx 1.15R_\times$ and later weakens as the wave expands to $r\gg\Rx$. The typical behavior of $\sigbg\propto r^{-3}$ in a neutron star magnetosphere gives $\gamma \propto r^{-4}$ at $r\gg\Rx$. 

%%%%%%%%%%% FIGURE %%%%%%%%%%%%%%%%%%
%%%%%%%%%%% FIGURE %%%%%%%%%%%%%%%%%%

\subsection{Shock dynamics}

The shock position $\xish(t)$ obeys an ordinary differential equation, which can be derived as follows. Suppose we are given an instantaneous position $\xish$ and wish to find $d\xish/dt=1-\bsh$, where $\bsh$ is the shock speed. This can be achieved in two steps: first determine the speed of the downstream plasma $\betad$ immediately behind the shock, and then find $\bsh$ from the shock jump conditions.

The downstream plasma velocity is determined by the known fields $\Ed$ and $\Bd$ in the undeformed part of the wave profile $E(\xi)$ behind the shock,
\beq
   \betad=\frac{\Ed}{\Bd}=\frac{\Ed}{\Bbg+\Ed}, \qquad \Ed=E_0\sin(\omega\xish). 
\eeq 
Using $E_0/\Bbg=r^2/2\Rx^2$, one finds the downstream compression factor $\cd$ (defined according to \Eq~(\ref{eq:kappa})):
\beq
\label{eq:cd}
   \cd^2=\frac{1+\betad}{1-\betad}=1+\frac{r^2}{\Rx^2}\sin(\omega\xish).
\eeq
Note that $\cd\ll 1$ (i.e. $\betad\approx -1$) if the jump of the electric field at the shock is small, $|(\Ed-\Ep)/\Ep| \ll 1$. The shock is very strong in the sense that $\gd\ll\gamma$, where $\gamma$ is the Lorentz factor of the upstream plasma crossing the shock.

Next, we use the shock jump condition to find $\bsh$. The shock Lorentz factor in the rest frame of the downstream plasma $\KFd$ is given by (see \citetalias{Beloborodov23}):
\beq
\label{eq:jump}
   \gshd\approx \frac{\cshd}{2}\approx (1+\chi)^{1/7}\sqrt{\sigma}\gg 1, \qquad \sigma\approx\frac{\sigbg}{2\gamma},
\eeq
where $\sigma$ is the magnetization parameter of the upstream flow (\Eq~\ref{eq:continuity}) with $\beta\approx -1$. The upstream Lorentz factor at the shock, $\gamma$, is controlled by plasma acceleration on the plateau (\Eq~\ref{eq:gamma}) and radiative losses induced by the precursor ahead of the shock (\Sect~\ref{precursor}); it can be significantly below $\gu$ given in \Eq~(\ref{eq:gu}), see Figure~\ref{fig:plateau}. The parameter $\chi$ appearing in the jump condition~(\ref{eq:jump}) is defined by
\beq
\label{eq:chi}
  \chi^{4/7} = \frac{\sT \Bbg }{\pi e}\,\gamma^2\cd^3.
\eeq

The jump condition~(\ref{eq:jump}) implies a relation between $\gamma$, $\cd$, and $\csh$. It is found using the transformation of the shock speed to the lab frame, which may be expressed as\footnote{Quantity $\c$ is defined in terms of any given $\beta$ in \Eq~(\ref{eq:kappa}). Relativistic addition of any velocities $\beta_1$ and $\beta_2$ is equivalent to multiplication of $\c_1$ and $\c_2$.}
 $\csh=\cshd\cd$. Then, we find
\beq
\label{eq:xish_diff}
   \frac{d\xish}{dt}=1-\bsh = \frac{2}{1+\csh^2}=\frac{2}{1+\mathring{\c}_{\rm sh}^2\cd^2}, 
\eeq
where $\cd$ and $\cshd$ are expressed in terms of $\xish$, $\gamma$, $r$, $\Bbg$, and $\sigbg$  (\Eqs~(\ref{eq:cd})-(\ref{eq:chi})). This gives the differential equation describing the shock motion $\xish(t)$. As shown below, the shock moves outward with a relativistic speed $\bsh\approx 1$, $\gsh\gg 1$. Therefore, the differential equation can be simplified to 
\beq
\label{eq:xish_diff1}
  \frac{d\xish}{dt}\approx\frac{1}{2\gsh^2}, \qquad \gsh\approx \gshd\cd.
\eeq

The shock is launched at $\xish\approx\xi_0=3\pi/2\omega$ at $t\approx \Rx/c$, and an initial phase of its evolution occurs with $\cd\ll 1$. At this phase, the differential equation is not needed, as $\xish(t)$ can be found in a simpler way. The condition $\cd\ll 1$ corresponds to a small jump of the electric field at the shock, $|\Delta E/\Ep|\ll 1$, and one can find $\xish$ neglecting $\Delta E$ and matching the unperturbed profile $E(\xi)=E_0\sin(\omega\xi)$ to $\Ep$: $E(\xish)=-\Bbg/2$. This gives 
\beq
\label{eq:xish_appr}
   \omega\xish\approx 2\pi -\arcsin\left(\frac{\Rx^2}{r^2}\right) \qquad \left(\cd\ll 1\right).
\eeq 
Note that $\xiin$ also satisfies $E(\xiin)=-\Bbg/2$, and so the plateau boundaries $\xiin$ and $\xish$ are symmetric about the point $\omega\xi_0=3\pi/2$ as long as $\cd\ll 1$. From \Eq~(\ref{eq:xish_appr}) we also find the shock Lorentz factor:
\beq 
\label{eq:gsh_appr}
   \gsh^2\approx \left(2\frac{d\xish}{dt}\right)^{-1}
   \approx \frac{\omega r }{4c} \sqrt{\frac{r^4}{\Rx^4}-1}  \quad\; \left(\cd\ll 1\right).
\eeq
This expression assumes $\gsh\gg 1$, which becomes satisfied almost immediately  after the magnetosonic wave crosses $\Rx$, since $\omega r/c\gg 1$.

The shock trajectory $\xish(t)$ follows \Eq~(\ref{eq:xish_appr}) only when $\cd\ll 1$. As $\cd$ grows with time, one has to switch to solving the differential equation~(\ref{eq:xish_diff1}). This is required before the shock reaches $\omega\xish=2\pi$ (at this point $\cd=1$). The deviation of $\xish(t)$ from \Eq~(\ref{eq:xish_appr}) is observed in the full MHD simulation \citepalias{Beloborodov23} and accurately reproduced by \Eq~(\ref{eq:xish_diff1}).

%##############################################################

\section{Precursor}
\label{precursor}

\subsection{Reference frames and notation}
\label{frames}

Besides the static lab frame $\KF$ (rest frame of the magnetar), our analysis of shock precursor emission will involve two other frames: the local rest frame of the upstream plasma flow $\tilde{\KF}$ and frame $\KFd$ where  the downstream plasma (immediately behind the shock) is at rest.

Parameters of the downstream fluid (at $\xi=\xish$) are denoted with subscript ``d.'' Subscript ``u'' will be used for the accelerated upstream flow at $\xi=\xi_0$, just before it begins to interact with the precursor. Parameters of the upstream flow interacting with the precursor and moving toward the shock (at $\xi_0<\xi<\xish$) will be denoted without subscript ``u.'' Accents will be added to symbols to indicate the reference frame in which the quantity is measured. For instance, $\mathring B$ is the magnetic field of the upstream flow measured in frame $\mathring\KF$, $B$ is the field measured in the lab frame $\KF$, and $\tB$ is the field measured in the local fluid frame $\tKF$.

The upstream plasma drifts relative to the lab frame with velocity $\bD=E/B\approx -1$, which corresponds to $\c\equiv\gD(1+\bD)\approx (2\gD)^{-1}\ll 1$.\footnote{Hereafter, we use subscript ``D'' for the fluid motion (the $\bE\times\bB$ drift) to distinguish it from the motion of individual particles, which are energized in the fluid frame by the precursor wave.} 
The upstream and downstream magnetic fields measured in the fluid rest frame are given by \Eq~(\ref{eq:compr}):
\beq
   \tBu=\c\Bbg,  \qquad
   \Bdd=\cd \Bbg.
\eeq

It is customary to view the shock and its precursor emission in the downstream frame $\KFd$. Therefore, it is useful to state the upstream Lorentz factor $\gDud$ and magnetic field $\Bud$ measured in frame $\KFd$. The upstream moves with $\bDud\approx -1$ and $\cud\equiv\gDud(1+\bDud)\approx(2\gDud)^{-1}\ll 1$. Velocity transformation from frame $\KF$ to frame $\KFd$ has the form $\mathring\c=\c/\cd$, which gives
\beq
   \gDud\approx \gD\cd.
\eeq
The upstream magnetic field measured in frame $\KFd$ is 
\beq
  \Bud=\gDud\tBu,
\eeq
as follows from field transformation $\Bud=\gDud(\tBu+\bDud\tE)$ with $\tE=0$ by definition of frame $\tKF$. The downstream field $\Bdd$ can be expressed in terms of $\Bud$ as follows
\beq
\label{eq:Bd}
  \Bdd=\frac{\cd}{\c}\tBu = \frac{\tBu}{\cud} = (1-\bDud)\Bud \approx 2\Bud.
\eeq

It is important to note that particles in the upstream plasma participate in two motions: fluid drift and oscillations in the fluid frame, which are driven by the precursor wave.\footnote{Particle motion in a strong radio wave, viewed in the local fluid frame $\tKF$ or static lab frame $\KF$, is discussed in Appendix~B in \citetalias{Beloborodov26}.} These oscillations are ultra-relativistic (as the precursor wave is strong), i.e. the particles have an average Lorentz factor $\tg\gg 1$ in the fluid frame $\tKF$. The particle Lorentz factors in frames $\KF$ and $\KFd$ are given by
\beq
\label{eq:gam_trans}
    \gamma=\gD\tg, \qquad \gud=\gDud\tg.
\eeq

Precursors of magnetized shocks are emitted by the plasma that enters the shock and begins gyration with Lorentz factor $\gud$ in the transverse magnetic field $\Bdd$. The precursor is described by its frequency $\ompred$ and energy density 
\beq
   \Upred=\frac{\langle \Epred^{\,2}\rangle}{4\pi}, 
\eeq
where $\Epred$ is the oscillating electric field of the precursor wave, and $\langle...\rangle$ signifies averaging over the oscillation period $2\pi/\ompred$.

\subsection{Precursor generation by non-radiative shocks}
\label{generation}

Shocks with negligible radiative losses have been studied in detail with fully kinetic simulations \citep{Iwamoto17,Plotnikov19,Sironi21,Vanthieghem25a,Bernardi25}. Plasma particles enter the shock and develop gyration with the Lorentz factor $\gud$, and the precursor wave emerges with frequency $\ompred$ comparable to the downstream Larmor frequency $\omd=e\Bdd/\gud\me c=2\,\tomB$, where
\beq
   \tomB=\frac{e\tBu}{\me c}. 
\eeq
Numerical simulations show that the precursor emission in frame $\KFd$ peaks at 
\beq
\label{eq:ompre_d}
   \ompred\approx 1.5\, \omd = 3\,\tomB = \frac{3e\Bud}{\gud \me c}.
\eeq
The emission is often called ``synchrotron maser.''

The precursor energy density is parameterized as
\beq
\label{eq:Epre}
   \Upred=\xiB\,\frac{\Bud^2}{4\pi}, \qquad \langle\Epred^{\,2}\rangle=\xiB \Bud^2,
\eeq
where $\xiB$ is a numerical factor (denoted as $\xi_B$ in \cite{Sironi21}). Numerical simulations show $\xiB\approx 10^{-2}$,  remarkably independent of the upstream magnetization $\sigma\gg 1$ and $\gud\gg1$. The precursor is generated because the shock transition from $\Bud$ to $\Bdd\approx 2\Bud$ is unsteady. The transition is supported by a layer of transverse electric current forming at the onset of the opposite gyration of $e^+$ and $e^-$ as they cross the shock. The gyration current is unstable on the Larmor timescale and its ``ringing" generates electromagnetic waves. 

The precursor waves have $\delta\bEud\perp\bBud$, similar to   magnetosonic waves. However, they are not MHD waves, because the precursor frequency measured in the upstream rest frame $\tKF$ far exceeds the plasma gyro-frequency $\tomB$. In contrast to MHD waves, the electromagnetic precursor  propagates faster than the shock, with the group speed practically equal to $c$ (slightly reduced due to the non-zero plasma density). The dimensionless strength parameter of the precursor is 
\beq
\label{eq:a}
   a \equiv \frac{e\langle\Epred^{\,2}\rangle^{1/2}}{\me c\, \ompred\; }
     \approx \frac{1}{3}\,\xiB^{1/2}\, \gud.
\eeq
It is defined here in frame $\KFd$, but the definition and value of $a$ are invariant under Lorentz boosts along the direction of wave propagation, so it is the same in the lab frame. Note that \Eqs~(\ref{eq:gam_trans}) and (\ref{eq:a}) imply
\beq
\label{eq:gDud}
  \gDud = \frac{\gud}{\sqrt{1+a^2}} \approx \frac{3}{\sqrt{\xiB}} \qquad (a\gg 1).
\eeq

The picture of precursor emission summarized above is changed if particles experience radiative losses before completing the shock transition, as occurs in monster shocks. Existing kinetic simulations of shocks did not include radiative losses and so far were limited to $\gud<10^3$; monster shocks reach much greater $\gud$.

Importantly, radiative losses can occur not only in the shock itself but also ahead of it, where the upstream plasma interacts with the precursor, which reduces $\gud$. Below we first discuss the losses inside the shock and then losses ahead of the shock.

\subsection{Radiative losses inside the shock}
\label{losses_shock}

The shock thickness is set by the Larmor radius $\gud \me c^2/e\Bdd$. The upstream particle with Lorentz factor $\gud$ develops gyration in the shock and experiences radiative losses. The losses inside the shock are significant if they occur on a timescale $\mathring{t}_{\rm loss}$ shorter than the Larmor time $\mathring{t}_{\rm L}$ (both measured in the downstream rest frame $\KFd$). Their ratio is \citepalias{Beloborodov23}
\beq
\label{eq:chish}
  \frac{\mathring{t}_{\rm L}}{\mathring{t}_{\rm loss}} \approx \frac{2\sT \Bud\,\gud^2}{\pi e}=  \chi^{4/7}.
\eeq
The parameter $\chi$ was defined in \Eq~(\ref{eq:chi}).
When $\chi\gg 1$, the particles end up executing the first full gyration in the shock with the ``radiation-reaction-limited'' Lorentz factor $\gud_{\rm rad}$ at which the Larmor timescale equals the cooling timescale,
\beq
   \gud_{\rm rad} \sim \left(\frac{\pi e}{2\sT \Bud}\right)^{1/2} \qquad (\chi\gg 1).
\eeq
Interpolating between the limits of $\chi\ll 1$ and $\chi\gg 1$, we can estimate that the Lorentz factor $\gud$ of the particle entering the shock is reduced by the factor of $(1+\chi)^{-2/7}$ before it fully develops downstream gyration. Then, the Larmor frequency in the shock is increased by the factor of $(1+\chi)^{-2/7}$, and one can expect an increased frequency of precursor emission:
\beq
\label{eq:ompre_d1}
  \ompred\approx 3\,\tomB (1+\chi)^{2/7} = \frac{3e\Bbg  (1+\chi)^{2/7}}{2\gamma \me c}.
\eeq
When $\chi\gg 1$, this gives 
\beq 
   \ompred\sim \frac{3e\Bud}{\gud_{\rm rad} \me c} 
        \sim 7\,\left(\frac{r_e}{c}\right)^{1/2}\left(\frac{e\Bud}{\me c}\right)^{3/2}
         \;\; (\chi\gg 1).
\eeq

Note that the shock parameter $\chi$ is sensitive to the upstream particle energy  $\gud \me c^2$. As shown below, the precursor waves significantly reduce $\gud$ by inducing radiative losses in the upstream plasma even before it reaches the shock. Then, $\chi\propto\gud^{7/2} $ can drop below unity, so that radiative losses inside the shock become unimportant.

\subsection{Upstream deceleration by the precursor}
\label{deceleration}

The upstream flow is strongly accelerated on the first half of the plateau ($\xiin<\xi<\xi_0$), as described in section~\ref{plateau} (Figure~\ref{fig:plateau}). In this zone, the flow is a cold MHD fluid ($\tg=1$) with the growing Lorentz factor $\gamma=\gD\approx (c\sigbg/r)(\xi-\xiin)$. The flow enters the second half of the plateau at $\xi_0$ with $\gD=\gu$ and magnetization $\sigu$ given by
\beq 
\label{eq:sigu}
   \gu=\! \frac{c\sigbg}{r}(\xi_0-\xiin), \quad  \sigu \!=\! \cu\sigbg \! \approx \frac{\sigbg}{2\gu}
   \quad (\xi=\xi_0).
\eeq
Then, before reaching the shock, the upstream flow interacts with the precursor, which occupies the second half of the plateau $\xi_0<\xi<\xish$ (the precursor was emitted in this region as the shock shifted from its initial position $\xi_0$ at $r=\Rx$ and the current $\xish(r)$). Interaction with the precursor increases $\tg$ and reduces $\gD=\gamma/\tg$. It also induces radiative losses that can strongly reduce $\gamma$.

This deceleration effect is controlled by the energy density of the precursor wave $\Upre$. Hereafter, we view it in the original lab frame $\KF$ (the rest frame of the unperturbed dipole magnetosphere). The downstream frame $\KFd$ used in the previous sections moves relative to the lab frame with speed $\betad$, and transformation of the precursor wave from $\KFd$ to $\KF$ is described by
\beq
\label{eq:prec_transf}
   \ompre =\cd\ompred,   \qquad  \Epre=\frac{\ompre}{\ompred}\,\Epred.
\eeq
The precursor energy density immediately ahead of the shock is\footnote{Remarkably, the upstream Lorentz factor $\gamma$ does not enter \Eq~(\ref{eq:Upre}): the emitted $\Upre$ depends only on the downstream motion $\cd=\gd(1+\betad)$, which is a simple function of time $t\approx r/c$ and shock position $\xish(t)$ in the magnetosonic pulse (\Eq~\ref{eq:cd}).} 
\beq
\label{eq:Upre}
   \Upre = \cd^2\,\Upred = \frac{\xiB \Bbg^2\cd^4}{16\pi}.
\eeq
The corresponding luminosity of the precursor (isotropic equivalent) is
\beq
\label{eq:Lpre}
    \Lpre(\xish)=4\pi r^2 c\,\Upre.
\eeq
As the shock moves with rate $d\xish/dt$, it adds precursor layers, forming the radio wave packet $\Lpre(\xi)$ in the region $\xi_0<\xi<\xish$. Assuming radial propagation of the emitted precursor wave, one finds that its density $\Upre(t,\xi)$ will decrease with time as $t^{-2}\propto r^{-2}$ 
while the precursor luminosity $\Lpre(\xi)$ remains constant at each $\xi$.
The precursor density at any $t$ and $\xi$ is then given by 
\beq
   U(t,\xi)=\frac{\Lpre(\xi)}{4\pi c r^2}=\frac{\Lpre(\xi)}{4\pi c^3 t^2}.
\eeq

The general problem of how a strong radio wave interacts with a magnetized $e^\pm$ plasma is described in \citetalias{Beloborodov26}. In the case of a monster shock precursor, the interaction occurs in the simple regime: 
\\
(1) The precursor wave drives regular oscillations of the plasma particles in the local fluid frame $\tKF$ (without activating irreversible stochastic heating, see section~\ref{stochastic} and Appendix~\ref{app:stochastic}). In this regime, the particle Lorentz factor (averaged over the oscillation) is $\tg=\sqrt{1+a^2}\approx a$.
\\
(2) The pattern of fluid motion in the precursor wave packet, $\c(\xi)=\gD(1+\bD)$, is quasi steady (Appendix~\ref{app:steady}). It  gradually evolves with time as the shock expands. 

The particle Lorentz factor measured in the lab frame is $\gamma=\gD\sqrt{1+a^2}$. Its quasi steady profile $\gamma(\xi)$ is affected by radiative losses, which may be written as $d\gamma/d\xi=-2\sT\Upre(\xi)\gamma^2/\me c$ (equivalent to \Eq~(51) in \citetalias{Beloborodov26}). As the plasma approaches the shock along the plateau, $\gamma$ changes with $\xi$ for two reasons: each particle gains energy with rate described by \Eq~(\ref{eq:gamma}) and loses energy by emitting radiation. Therefore, the profile of $\gamma(\xi)$ ahead of monster shocks obeys the equation 
\beq
\label{eq:dyn}
  \frac{d\gamma}{d\xi} =  \frac{c\sigbg}{r} - \frac{2\sT\Upre\gamma^2}{\me c}.
\eeq

The radiative loss term may be thought of as inverse Compton scattering of the radio wave. It can be derived as follows. The losses are caused by the fast-oscillating acceleration of the particle in the electromagnetic field of the radio wave. Note that the precursor wave strongly dominates over the MHD electromagnetic field $\tB$ when viewed in the local fluid frame $\tKF$:
\beq
  \frac{\tEpre}{\tB}\approx 2\gDud^2\frac{\Epred}{\Bud}\approx 2\left(\frac{\gud}{a}\right)^2\eps^{1/2}\approx \frac{18}{\eps^{1/2}}\gg 1.
\eeq
The last equality has used \Eq~(\ref{eq:a}) at the shock to show $\tEpre/\tB$ at $\xi=\xish$; however, $\tEpre\gg\tB$ holds at all $\xi$ across the precursor due to the large $\gDud(\xi)$. Then, each oscillating particle with an instantaneous four-velocity $u^\mu$ radiates the Lorentz-invariant power (see Appendix~D in \citetalias{Beloborodov26}):
\beq
\label{eq:dEe}
  \dEe\approx \frac{c\sT}{4\pi}u_\xi^2(\Epre)^2, \qquad u_\xi\equiv -u^\mu l_\mu,
\eeq
where $l^\mu=(1,\boldsymbol{n})$ is the wave direction in spacetime. While $u^\mu$ strongly oscillates in the radio wave, $u_\xi$ stays nearly constant during each oscillation \citepalias{Beloborodov26}. One  then finds the emitted power averaged over the oscillation: $\langle\dEe\rangle=c\sT u_\xi^2\Upre$. This expression is Lorentz invariant and can be used in any frame moving along $\pm\bn$. In the lab frame, the upstream plasma in front of the monster shock flows relativistically opposite to $\bn$, and each particle has $u_\xi\approx 2\gamma$. This gives $d\gamma/dt=-\langle\dEe\rangle/\me c^2=-4c\sT\Upre\gamma^2$. Then, using $d\xi/dt=1-\bD\approx 2$, one obtains the radiative loss term stated in \Eq~(\ref{eq:dyn}).

%############################################################################

\section{Numerical solution}
\label{numerical}

\subsection{Evolution of the shock+precursor system}

The equations governing the coupled evolution of the shock and its precursor can be summarized as follows.
\\
\\
(i) The shock trajectory $\xish(t)$ obeys \Eq~(\ref{eq:xish_diff1}):
\beq
\label{eq:xish_diff2}
  \frac{d\xish}{dt} \approx\frac{\gamma}{\sigbg \cd^2 (1+\chi)^{2/7}}, 
   \quad \chi^{4/7} = \frac{\sT \Bbg }{\pi e}\,\gamma^2\cd^3,
\eeq
where $t\approx r/c$ and $\cd^2(r,\xish)$ is given by \Eq~(\ref{eq:cd}). The evolution of $\xish$ is not immediately found from \Eq~(\ref{eq:xish_diff2}) alone, since it involves the Lorentz factor $\gamma$ of the upstream plasma particles entering the shock.
\\
\\
(ii) At any given time $t$ (when the shock has radius $r=ct$), the profile of $\gamma(\xi)$ and its value at the shock $\gamma(\xish)$ are found by integrating \Eq~(\ref{eq:dyn}), which can be rewritten as
\beq
\label{eq:energy1}
    \frac{d\gamma}{d\xi}= \frac{c\sigbg}{r} - \frac{\sT\Lpre(\xi)}{2\pi \me c^2 r^2} \gamma^2
    \qquad (\xi_0<\xi<\xish).
\eeq
One finds $\gamma(\xi)$ by integrating this equation across the small precursor width $c(\xish-\xi_0)\ll r$, so $r$ and $\sigbg(r)$ can be taken as constant in the integral. The boundary condition $\gamma(\xi_0)=\gu$ is given in \Eq~(\ref{eq:sigu}). The result for $\gamma(\xi)$ is determined by the known precursor luminosity $\Lpre(\xi)$, which was produced at earlier times.
\\
\\
(iii) The production of precursor luminosity $\Lpre(\xi)$ is described by \Eq~(\ref{eq:Lpre}). At each time step $\delta t$ a new layer $\delta \xi=(d\xish/dt)\delta t$ is added to the precursor wave packet. The luminosity of the added layer $\delta\xi$ is 
\beq
\label{eq:Lpre2}
    \Lpre=\frac{1}{4}\,\xiB\, c\,r^2 \Bbg^2\cd^4.
\eeq

\Eqs~(\ref{eq:xish_diff2}), (\ref{eq:energy1}), and (\ref{eq:Lpre2}), together with \Eq~(\ref{eq:cd}) for $\cd$,
form a closed system describing the evolution of shock+precursor. Note that it does not involve $a$ or $\ompre$. The system can be easily evolved numerically. The evolution can also be described analytically, as done in section~\ref{analytical} below. 

Once $\xish(t)$ and $\gamma(t)$ are obtained, one can also evaluate the frequency of the precursor wave using \Eqs~(\ref{eq:ompre_d1}) and (\ref{eq:prec_transf}):
\beq
\label{eq:ompre1}
    \ompre(\xi)\approx \frac{3\omB(1+\chi)^{2/7}\cd}{2\gamma}, \qquad \omB=\frac{e\Bbg}{\me c}.
\eeq
Except for a negligible (short) initial phase, the correction factor $(1+\chi)^{2/7}$ equals unity as shown below, so it can be omitted in \Eq~(\ref{eq:ompre1}).

The coordinate $\xi$ equals the observer time measured from the beginning of the magnetosonic pulse. It is convenient to define 
\beq
   \tobs=\xi-\xi_0, 
\eeq
so that $\tobs=0$ corresponds to the beginning of the precursor emission. The calculated $\Lpre(\xi)$ and $\ompre(\xi)$ define the predicted FRB, $\Lpre(\tobs)$ and $\ompre(\tobs)$

\subsection{Parameters of the problem}

The shock problem formulated above is completely specified by four parameters. Two parameters describe the background magnetosphere: $\mu=r^3\Bbg$ and $\N=r^3\nbg$. Magnetars typically have a magnetic dipole moment $\mu$ between $10^{32}$ and $10^{33}$\,G\,cm$^3$. Their density parameter $\N$ was previously estimated as $\N\sim 10^{37}$ \citep{Beloborodov20}, so we will use it as a fiducial value. Two other parameters describe the magnetospheric perturbation: frequency $\nu=\omega/2\pi$ and power $L$. The characteristic frequency is in the kHz band, so our sample models will have $\nu=1$\,kHz and 10\,kHz. The power of magnetospheric perturbation $L$ can vary in a broad range. Most of the perturbation energy is radiated in X-rays, so $L$ may be estimated from X-ray observations. The routinely observed X-ray bursts from magnetars have luminosities between $10^{39}$ and $10^{43}$\,erg/s.\footnote{Magnetars also produce giant flares with luminosities up to $10^{47}$\,erg\,s$^{-1}$. Such super-strong perturbations generate monster shocks of highest power. However, their radio emission efficiency $\Lpre/L$ is tiny inside the magnetosphere, as shown below. When shocks expand far beyond the light cylinder, they become capable of emitting bright cosmological FRBs \citep{Beloborodov20}. Here, we focus on shocks {\it inside} the magnetosphere, which may explain the relatively weak FRBs from SGR~1935+2154.}   

These parameters determine the radius of shock formation $\Rx$ (\Eq~\ref{eq:Rx}). Note that rotation of the magnetar does not affect the results until the shock expands to the light cylinder $\RLC=cP/2\pi\approx 1.4\times 10^{10} (P/3\,{\rm s})$\,cm, where we normalized the rotation period $P$ to 3\,s, close to $P=3.25$\,s observed in SGR~1935+2154. As shown below, radio emission from the expanding shock peaks at $r\approx 10^9{\rm \,cm}<\RLC$.

We will present detailed results for two sample models:
\\
{\bf Model~S:} a relatively strong perturbation with $L=10^{43}$\,erg/s. It is similar to the monster shock calculated in \citetalias{Beloborodov23}, with $\mu=10^{33}\,$G\,cm$^3$ and $\nu=10\,$kHz, but now including the precursor effect on the shock dynamics.
\\
{\bf Model~W:} a weaker perturbation, $L=10^{41}\,$erg/s. It is chosen to represent the activity of SGR~1935+2154, with $\mu=2\times 10^{32}$\,G\,cm$^3$ and $\nu=1$\,kHz.

\subsection{Results}

%%%%%%%%%%% FIGURE %%%%%%%%%%%%%%%%%%
\begin{figure*}[t]
% \hspace*{-0.9cm}
\includegraphics[width=0.48\textwidth]{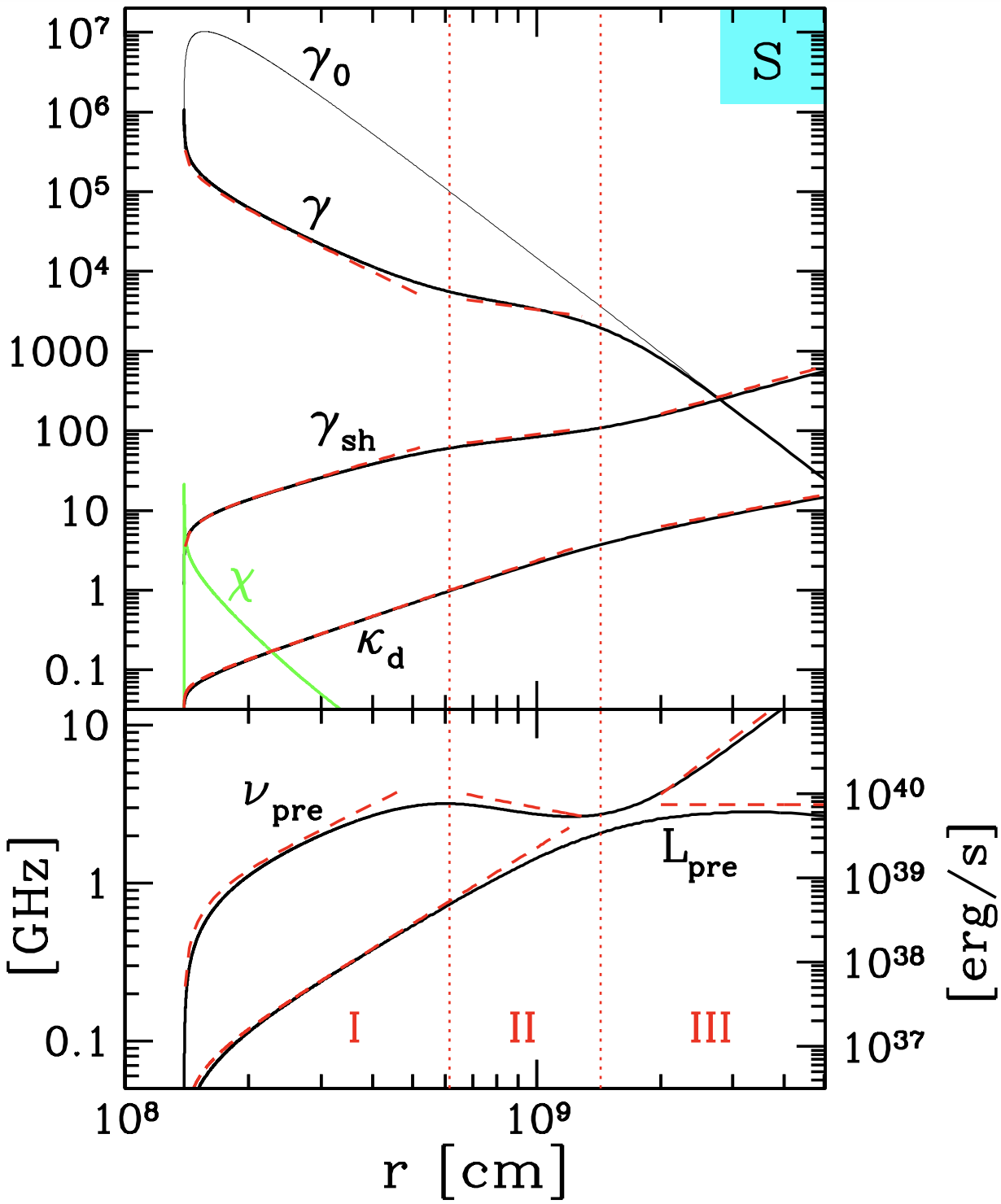} 
\hspace*{5mm}
\includegraphics[width=0.48\textwidth]{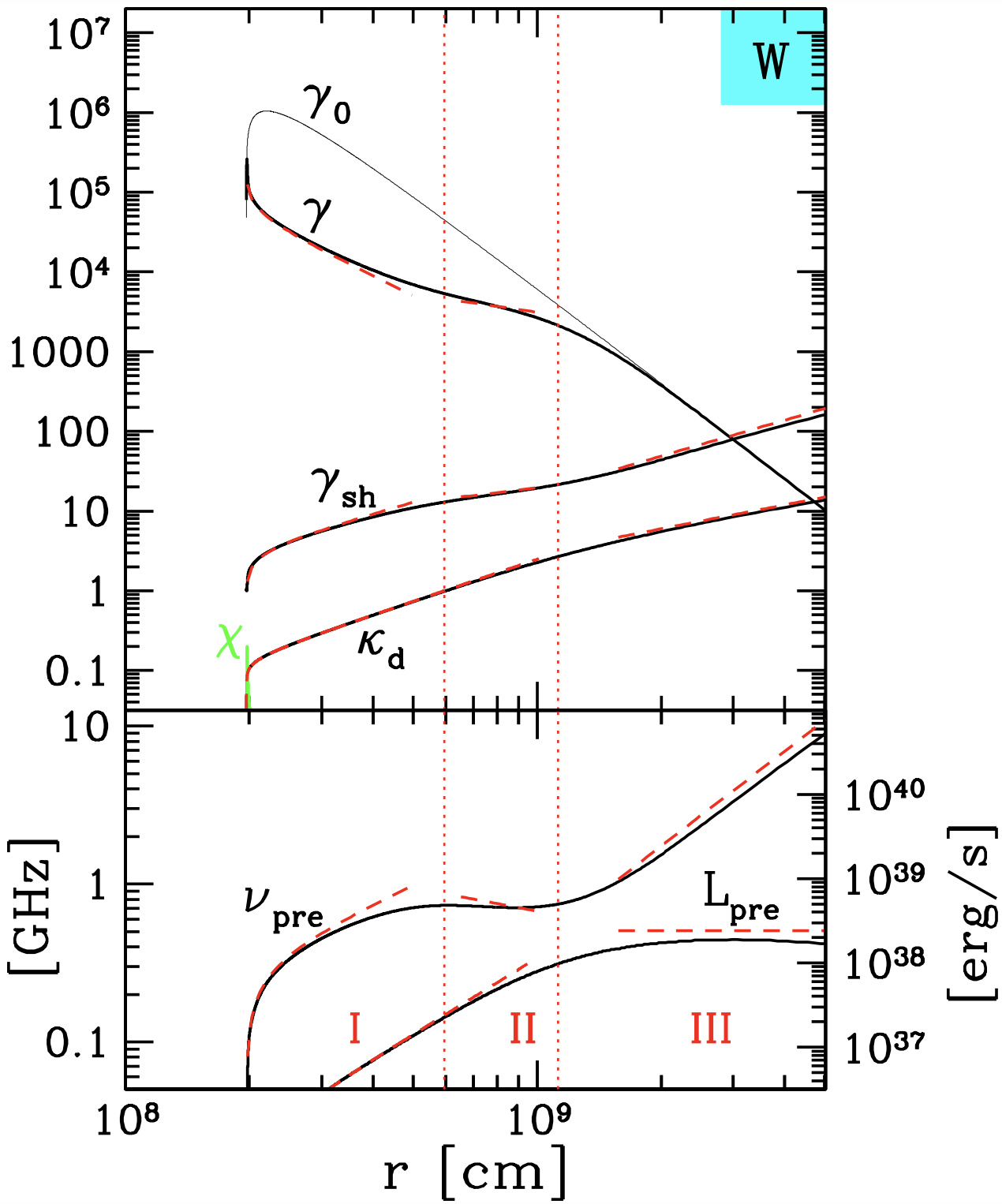} 
\caption{
Evolution of the shock parameters with radius $r=ct$ (starting from $\Rx$) in Model~S (left) and Model~W (right). Thick solid curves in the upper panels show the Lorentz factor of plasma particles entering the shock $\gamma$, Lorentz factor of the shock itself $\gsh$, and $\cd\equiv\gd(1+\betad)$ that describes the downstream fluid motion immediately behind the shock; the green curve shows the parameter $\chi$. Thin solid curve shows $\gamma_0(r)$; it represents the value of $\gamma$ found when the precursor effect on the upstream flow is neglected. The lower panels present the luminosity $\Lpre$ and frequency $\nupre$ of the precursor emission produced by the shock. In all panels, dashed red curves show the analytical results presented in \Sect~\ref{analytical} and derived in Appendix~\ref{derivation}. The vertical red dotted lines indicate the boundaries between phases I, II, and III.}
\label{fig:sim}
\end{figure*}
%%%%%%%%%%% FIGURE %%%%%%%%%%%%%%%%%%

First, as a test, we have calculated the shock evolution neglecting the precursor. This is done by setting the precursor luminosity $\Lpre=0$ and solving the remaining two equations for the shock position $\xish$ and upstream Lorentz factor $\gamma$. Then, Model~S should reproduce the results shown in Figure~5 of  \citetalias{Beloborodov23}, but now with a much simpler approximate calculation. To highlight the neglect of the precursor, the result for $\gamma(r)$ is denoted as $\gamma_0(r)$. It is given by
\beq 
\label{eq:gamma0}
   \gamma_0=\frac{\Wp\sigbg}{r},
\eeq
where $\Wp=c(\xish-\xiin)$ (\Sect~\ref{plateau}). We have verified that both $\xish(r)$ and $\gamma_0(r)$ obtained in our simplified model are nearly indistinguishable from the full MHD simulation performed in  \citetalias{Beloborodov23}. So, the dynamics of monster shocks can be accurately calculated without expensive MHD simulations. 

Next, we include the precursor effect, with the self-consistent calculation of luminosity $\Lpre$. The results for Models~S and W are presented in Figure~\ref{fig:sim}. One can see that the precursor strongly impacts the shock dynamics: radiative losses induced by the precursor steal a large part of energy gained by the plasma from the kHz magnetosonic wave, so the upstream particles come to the shock with a reduced Lorentz factor $\gamma<\gamma_0$ (a snapshot of Model~W is shown Figure~\ref{fig:plateau}). The losses also affect other shock parameters $\gsh$, $\cd$, and $\chi$. In particular, $\chi\propto\gamma^{7/2}$ (\Eq~\ref{eq:chi}) is strongly reduced compared to the model with neglected precursor.

Figure~\ref{fig:sim} also shows the evolution of luminosity $\Lpre(r)$ and frequency $\nupre(r)$ of the precursor emission as the shock radius $r$ grows. In both presented models, the precursor energy peaks in the GHz band. This a general feature of monster shocks, as explained analytically in \Sect~\ref{analytical} below. In both models, the luminosity $\Lpre$ grows with $r$ and then saturates at $r>10^9$\,cm. 

%%%%%%%%%%% FIGURE %%%%%%%%%%%%%%%%%%
\begin{figure}[t]
% \vspace*{-1.7mm}
\includegraphics[width=0.45\textwidth]{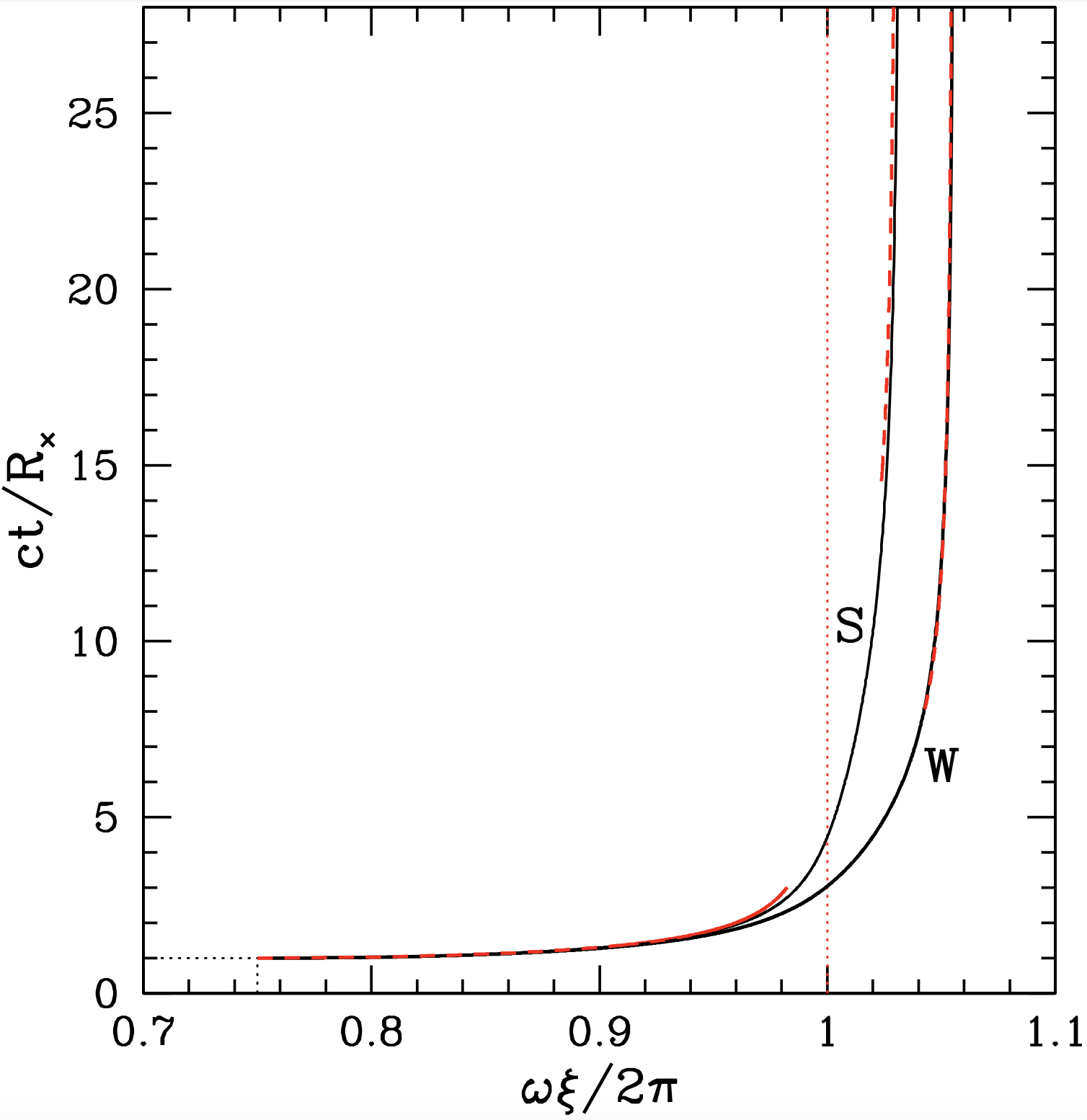} 
\caption{Shock trajectory on the $\xi$-$r$ plane, where $r=ct$ is the shock radius. The shock forms at $\xi_0=3\pi/2\omega$ and $r=\Rx$ (this moment is indicated by the black dotted lines). The observer time (measured from the moment of shock formation $t_0$) is $\tobs=\xish-\xi_0$. Two simulations are presented in the figure: Model~S and Model~W. Red dashed curves show the analytical approximation described in Appendix~\ref{derivation}. Vertical red dotted line indicates the moment when the shock crosses radius $R_1$ ($\cd=1$), entering phase~II of its evolution.}
\label{fig:xish}
\end{figure}
%%%%%%%%%%% FIGURE %%%%%%%%%%%%%%%%%%

Although the shock sustains the highest $\Lpre$ as it continues to propagate at $r\gg 10^9$\,cm, the contribution of this extended zone to the burst energy $\EFRB=\int \Lpre\, d\xi$ is small because at large $r$ the shock stalls in the $\xi$ coordinate (and so also stalls in the observer time $\tobs$). Thus, the highest luminosity occurs in a small interval $\delta\tobs$ in the end of the observed burst. This effect is caused by the growing Lorentz factor of the shock $\gsh(r)$, which determines $d\xish/dt\approx (2\gsh^2)^{-1}$ (\Eq~\ref{eq:xish_diff1}). The shock motion in $\xi$ is shown in Figure~\ref{fig:xish}. As one can see in the figure, $\xish(r)$ practically freezes at radii $r>2\times 10^9$\,cm, reaching $\omega\xi_\star/2\pi \approx 1.025$ in Model~S and 1.055 in Model~W. 

%%%%%%%%%%% FIGURE %%%%%%%%%%%%%%%%%%
\begin{figure}[t]
\includegraphics[width=0.45\textwidth]{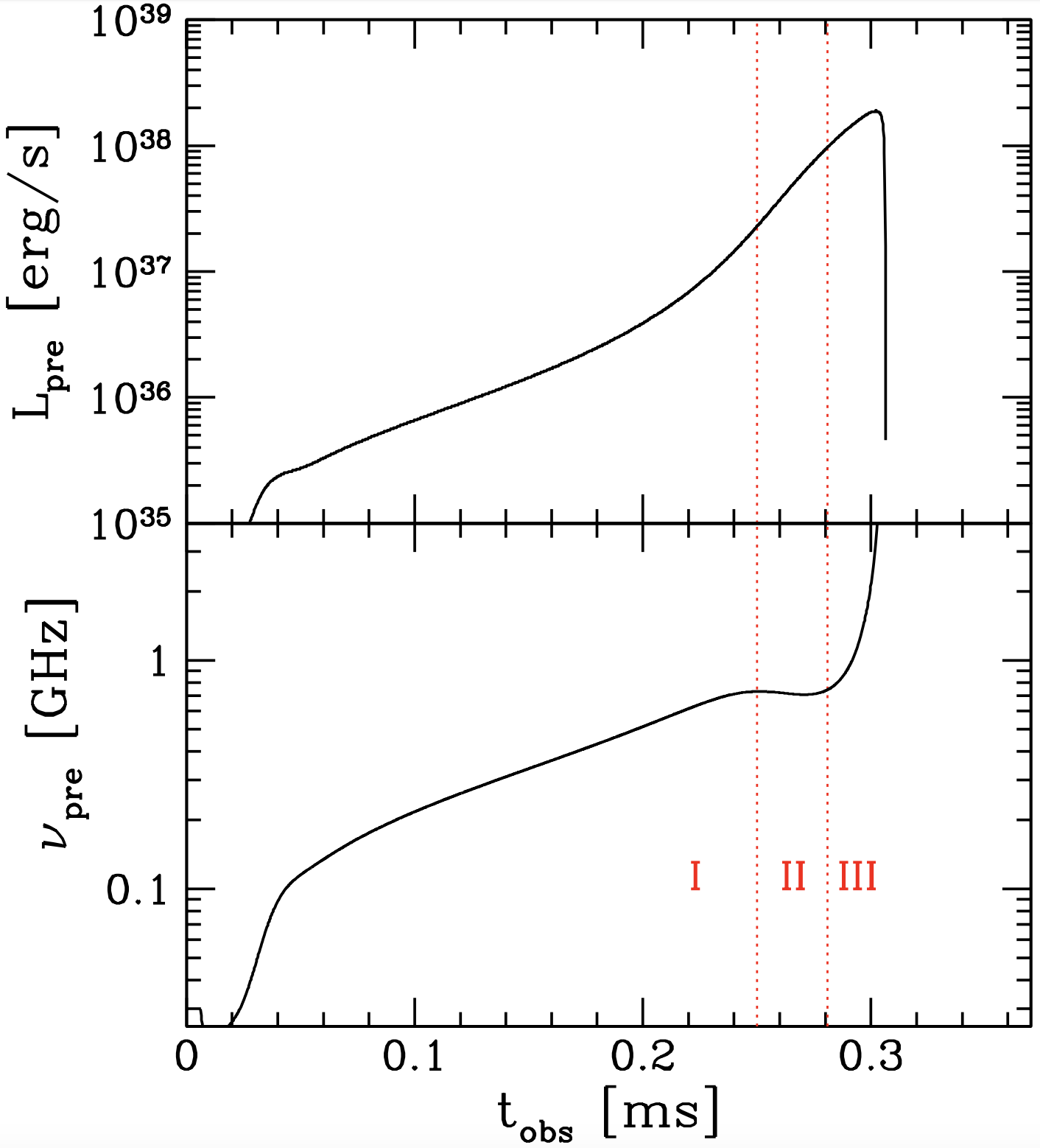} 
\caption{Radio burst predicted in Model~W: the burst light curve $L_{\rm pre}(\tobs)$ (upper panel) and the evolution of the burst frequency $\nu_{\rm pre}=\ompre/2\pi$ (lower panel). The red dotted lines indicate the boundaries between phases I, II, and III.}
\label{fig:obsW}
\end{figure}
%%%%%%%%%%% FIGURE %%%%%%%%%%%%%%%%%%
%%%%%%%%%%% FIGURE %%%%%%%%%%%%%%%%%%
\begin{figure}[t]
\includegraphics[width=0.45\textwidth]{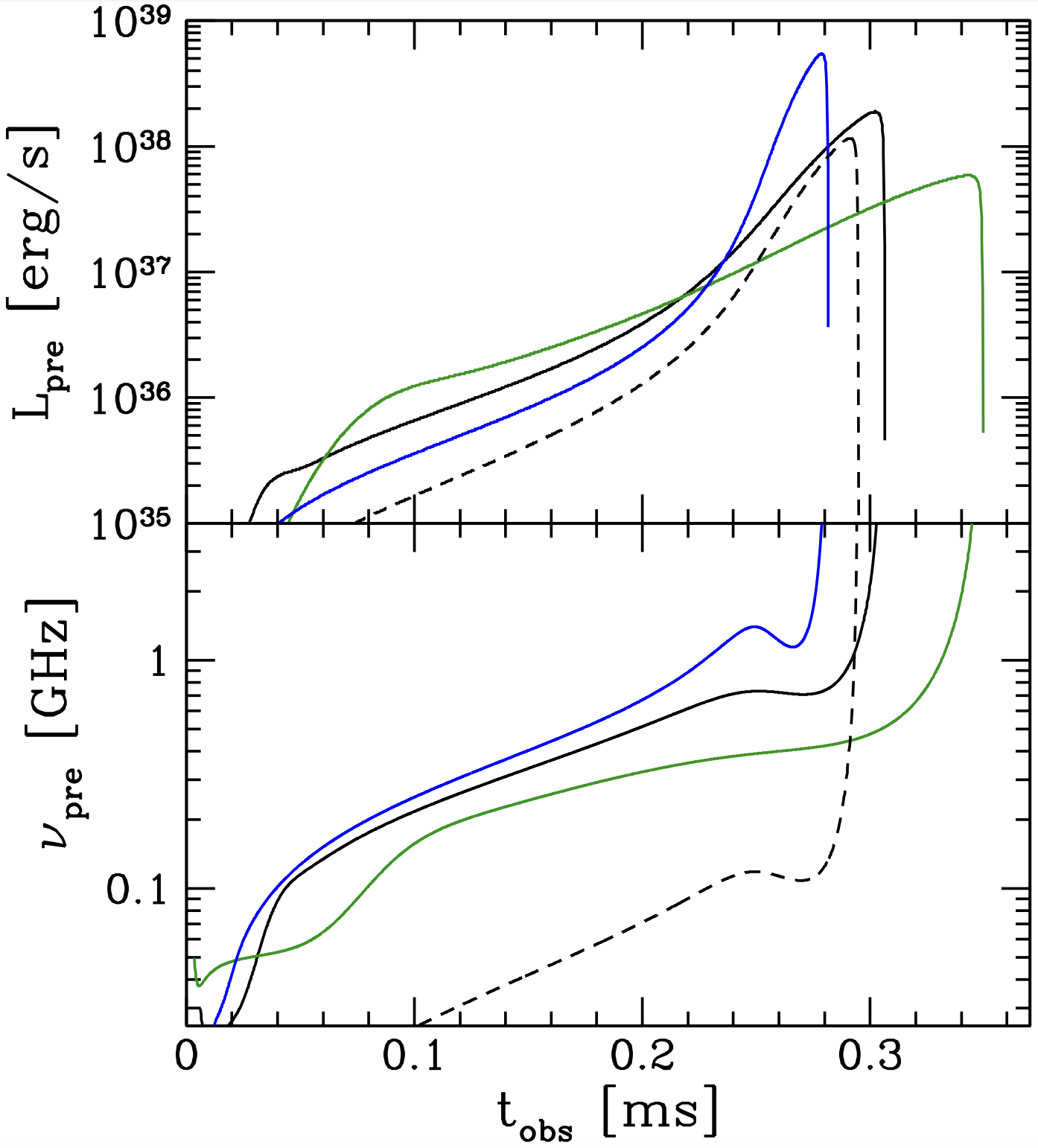} 
\caption{FRBs expected from monster shocks. Solid black curves show our fiducial Model~W with $L=10^{41}$\,erg\,s$^{-1}$ and $\N=10^{37}$. Colored curves shows  models with $L=10^{40}$\,erg\,s$^{-1}$ (green) and $L=10^{42}$\,erg\,s$^{-1}$ (blue), keeping $\N$ the same. Dashed black curves show the model with $L=10^{41}$\,erg\,s$^{-1}$ and $\N=10^{36}$. In all the four models, the magnetar has $\mu=2\times 10^{32}$\,G\,cm$^3$ and the magnetosonic perturbation that launches the shock has frequency $\nu=1$\,kHz.}
\label{fig:obs}
\end{figure}
%%%%%%%%%%% FIGURE %%%%%%%%%%%%%%%%%%

%%%%%%%%%%% FIGURE %%%%%%%%%%%%%%%%%%
\begin{figure}[t]
\includegraphics[width=0.47\textwidth]{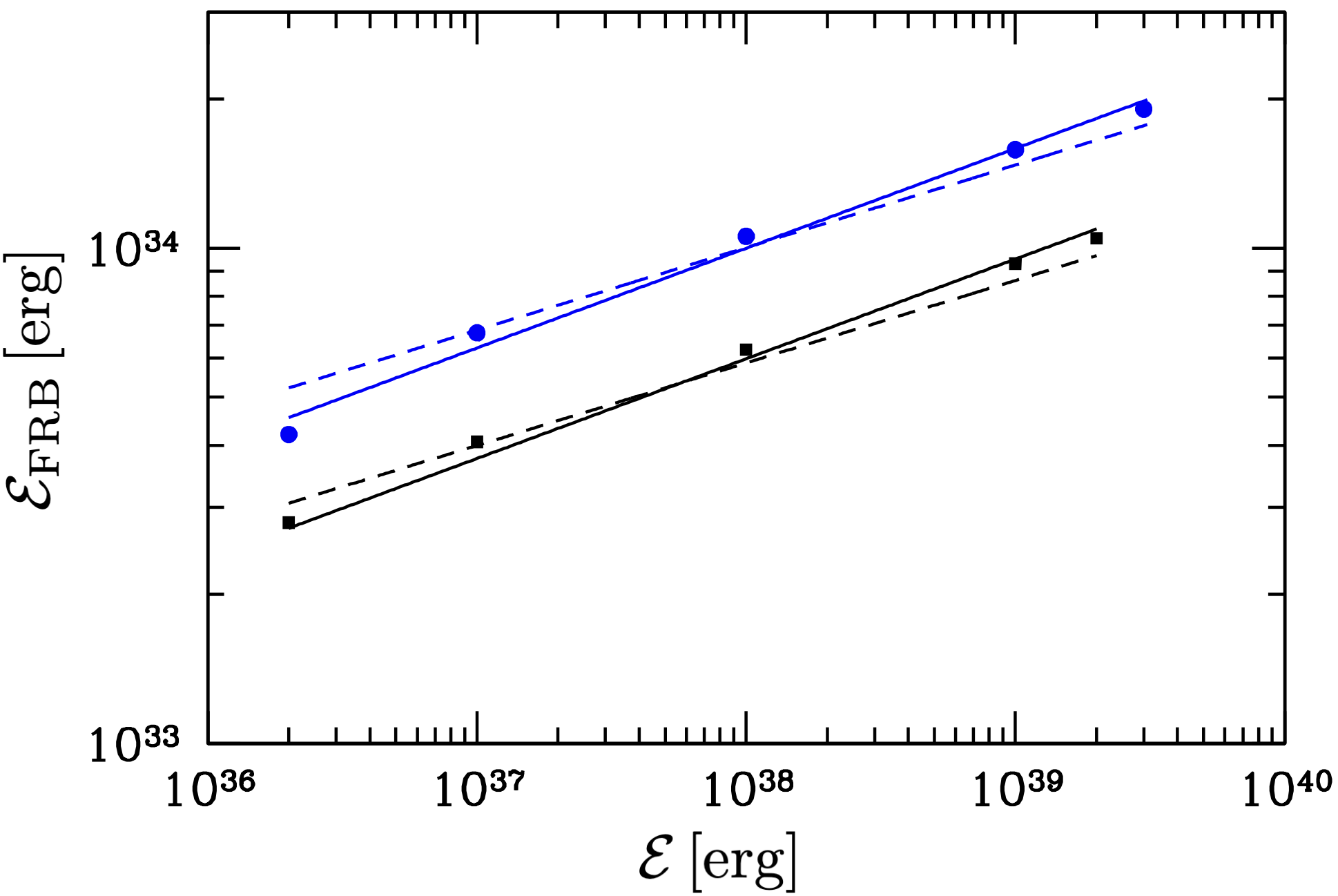} 
\caption{Energies of FRBs emitted by monster shocks from magnetars with $\N=10^{37}$ and $\mu=2\times 10^{32}$\,G\,cm$^3$ (black squares) and $10^{33}$\,G\,cm$^3$ (blue circles). For each $\mu$, the figure shows five numerical models with different energies $\E$ of the kHz magnetosonic perturbation that launched the shock. The solid and dashed lines show the analytical approximations given by \Eqs~(\ref{eq:Enpre1}) and (\ref{eq:Enpre2}), respectively.}
\label{fig:Enpre}
\end{figure}
%%%%%%%%%%% FIGURE %%%%%%%%%%%%%%%%%%

The value of $\xi_\star$ determines the observed duration of the radio burst: $T_{\rm FRB}=\xi_\star-\xi_0$. In particular, in Model~W, we find $T_{\rm FRB}\approx 0.61\pi/\omega\approx 0.3$\,ms. The predicted time profiles of the burst luminosity $\Lpre(\tobs)$ and frequency $\ompre(\tobs)$ are shown in Figure~\ref{fig:obsW}. The luminosity has a narrow peak toward the end of the burst within a short time interval $\delta \tobs\sim 0.05$\,ms. 

Additional models of FRB emission from monster shocks with different parameters are shown in  Figure~\ref{fig:obs}. The emission frequency $\nupre$ remains in the GHz band and weakly depends on the parameters $\mu$, $L$, and $\nu$; it is most sensitive to the density parameter $\N$. The total energies of predicted FRBs are shown for two sequences of models in Figure~\ref{fig:Enpre}. They vary around $\EFRB\sim 10^{34}$\,erg across the broad range of magnetosonic perturbation power $10^{-2}<L_{41}<10^2$. 
The production of $\EFRB$ peaks at $r\sim 10^9$\,cm in all models. 
The next section gives an analytical explanation for these numerical results.

%########################################################################

\newpage

\section{Analytical solution}
\label{analytical}

\subsection{Shock dynamics}

In Appendix~\ref{derivation} we derive an approximate analytical solution for the shock evolution during three expansion phases: $r<R_1$ (phase~I), $R_1<r<R_{\rm rad}$ (phase~II), and $r>R_{\rm rad}$ (phase~III). The results are shown by the dashed red lines in Figure~\ref{fig:sim}. In particular, $\cd\equiv\gd(1+\betad)$ and $\gamma$ are given by

\begin{eqnarray}
\label{eq:cd_an}
  \cd^4\approx  \left\{\begin{array}{lr}
    \displaystyle{ \left( \frac{r}{R_1} \right)^{7} \left(1-\frac{\Rx^4}{r^4}  \right)^{1/2} } & 
       \vspace*{2mm}
      \;\; \Rx<r\ll \Rrad \\
    \displaystyle{ \frac{5\pi}{4} \frac{r^4}{\Rx^2\Rrad^2}  } & \quad r\gg \Rrad
                            \end{array}\right. \quad
\end{eqnarray}

\begin{eqnarray}
\label{eq:g_an}
  \gamma \approx \left\{\begin{array}{lr}
    \displaystyle{ \frac{r}{\cd^2}\sqrt{\frac{2}{\sT\N\xiB}} }  
        \vspace*{2mm}
    &  \quad\; \Rx<r< R_1 \\
  \displaystyle{ \frac{2 r}{\cd \sqrt{\sT \N \xiB}}  }  
           \vspace*{1mm}
   &  \quad R_1< r <\Rrad \\
    \displaystyle{ \frac{\mu^2}{4\me c \N \omega \, r^4} }
   & r \gg \Rrad
                            \end{array}\right. 
\end{eqnarray}
We have neglected here the correction due to $\chi\neq 0$, since it is significant only in the small region of $r\approx\Rx$ (Figure~\ref{fig:sim}). Note also that the precise shock formation radius $r_c$ is slightly larger than $\Rx$ \citepalias{Beloborodov23}, which corresponds to the initial (minimum) value of $1-\Rx^4/r^4$:
\beq
\label{eq:min}
  \min \left(1-\frac{\Rx^4}{r^4}\right)\approx \sqrt{\frac{3}{2}}\,\frac{c\,\omega}{\Rx\sigx}.
\eeq

The shock Lorentz factor $\gsh$ can be expressed in terms of $\cd$ and $\gamma$ using the shock jump condition. This condition is stated in \Eq~(\ref{eq:jump}) in the downstream frame $\KFd$: $\gshd\approx (\c\sigbg)^{1/2}$ [omitting the factor $(1+\chi)^{2/7}\approx 1$]. Transformation from frame $\KFd$ to the lab frame is given by $\mathring\c_{\rm sh}=\c_{\rm sh}/\cd$. Then, using $\mathring\c_{\rm sh}\gg 1$ and $\c_{\rm sh}\gg 1$), one finds $\gsh=\gshd\cd$, and hence
\beq
\label{eq:gsh_jump}
   \gsh^2\approx \cd^2\c\sigbg \approx \frac{\sigbg \cd^2}{2\gamma}.
\eeq
This expression, combined with \Eqs~(\ref{eq:cd_an}) and (\ref{eq:g_an}), gives the analytical approximation for $\gsh(r)$ in all three phases I, II, and III of shock expansion. Note that the shock moves outward with $\gsh\gg 1$. Its Lorentz factor becomes particularly large during phase~III: 
$\gsh^2\approx (\sqrt{5\pi}/2)(\nu r^3/c\Rx\Rrad)$ at $r\gg\Rrad$ (where $\nu=\omega/2\pi$).

The boundaries between the three evolution phases, $R_1$ and $\Rrad$, are given by
\beq
\label{eq:xrad}
  \frac{R_1}{\Rx} \approx \left(\frac{2^{9/2} \SS}{\pi^2}\right)^{1/14}, \quad 
  \frac{\Rrad}{\Rx}=\left(\sqrt{\frac{5\pi}{2}}\, \SS \right)^{1/9},
\eeq
\beq
\label{eq:psi}
   \SS\equiv \frac{\sT\xiB L^{5/2}}{\me^2 c^{9/2}\mu\N\omega^2}\approx 4.58\times 10^5\,\frac{L_{41}^{5/2}\xiB_{-2}}{\mu_{33}\N_{37}\nu_3^2},
\eeq
or
\beq
\label{eq:R1}
   R_1=1.18\times 10^9\,\frac{\mu_{33}^{3/7}\xiB_{-2}^{1/14}}{\N_{37}^{1/14} L_{41}^{1/14}\nu_3^{1/7}} {\rm ~cm},
\eeq
\beq
\label{eq:Rrad}
  \Rrad=2.1\times 10^9\,\frac{\mu_{33}^{7/18} L_{41}^{1/36} \xiB_{-2}^{1/9}}{\N_{37}^{1/9} \nu_3^{2/9}}  {\rm ~cm}.
\eeq

The analytical description summarized above may be used when $\SS\gg 1$ and $c/\omega\Rx\gg 1$. This parameter space roughly corresponds to the power $L$ of kHz magnetosonic perturbations in the range of $10^{-2}<L_{41}<10^2$.

\subsection{Parameters of the predicted radio burst}

The energy of the produced radio burst is 
\beq
   \Enpre=\int \Lpre\, d\tobs = \int \Lpre\, d\xish = \int  \frac{\Lpre\,dr}{2\gsh^2 c}.
\eeq
Using \Eqs~(\ref{eq:Lpre2}) for $\Lpre$ and \Eq~(\ref{eq:gsh_jump}) for $\gsh^2$, we obtain
\beq
   \Enpre =  \int  \pi \me c^2 \N  \xiB\,  \gamma\, \cd^2\, \frac{dr}{r}.
\eeq
Note that $\gamma\cd^2\propto r^{11/4}$ during phase~II and $\gamma\cd^2\propto r^{-2}$ during phase~III (see \Eqs~(\ref{eq:cd_an}) and (\ref{eq:g_an})). The integral peaks at $\Rrad$, and can be written as 
\beq
   \Enpre \approx  \alpha\, \pi \me c^2 \N  \xiB\,  (\gamma\, \cd^2)_{\Rrad},
\eeq
with the numerical coefficient  $\alpha\approx 1$: the contributions from phases~II and III, $\alpha_{\rm II}\approx 4/11$ and $\alpha_{\rm III}\approx 1/2$, plus a small contribution from phase~I add up to $\alpha\approx 1$. The value of $\gamma\cd^2$ at $\Rrad$ can be evaluated by using the solution for phase~II (this gives somewhat better accuracy than using the steeper solution in phase~III) and substituting $\xrad$ (\Eq~\ref{eq:xrad}). This gives
\beq
\label{eq:Enpre1}
  \Enpre\approx 10^{34}\, \frac{\E_{38}^{29/144} \mu_{33}^{23/72} \N_{37}^{23/72} \xiB_{-2}^{49/72}}{\nu_3^{23/144}} {\rm~erg}.
\eeq
Where we defined the characteristic energy of the kHz disturbance that launched the shock as $\E=L/\nu$.

An alternative simple approximation for $(\gamma\cd^2)_{\Rrad}$ uses the extrapolation of $\gamma\cd^2$ from phase~III, which yields
\beq
   \Enpre\approx \alpha\frac{\sqrt{5\pi}}{2}\,\frac{\eps L}{\nu}\frac{\Rx^3}{\Rrad^3}\approx\alpha\frac{(5\pi)^{1/3}}{2^{5/6}}\,\frac{\eps L}{\nu\SS^{1/3}}.
\eeq
This expression gives a suitable approximation with $\alpha=0.55$ instead of 1:
\beq
\label{eq:Enpre2}
    \Enpre\approx 10^{34}\, \frac{\E_{38}^{1/6}\mu_{33}^{1/3}\N_{37}^{1/3}\xiB_{-2}^{2/3}}{\nu_3^{1/6}}\,{\rm erg}.
\eeq
\Eqs~(\ref{eq:Enpre1}) and (\ref{eq:Enpre2}) both give good approximations for $\Enpre$ computed from numerical models (Figure~\ref{fig:Enpre}). 

Next, consider the luminosity $\Lpre$ of the predicted radio burst. It is given by \Eq~(\ref{eq:Lpre2}). Substituting the analytical solution for $\cd$ (\Eq~\ref{eq:cd_an}) we find
\begin{eqnarray}
\label{eq:Lpre_an}
  \Lpre \approx \left\{\begin{array}{lr}
           \vspace*{1mm}
  \displaystyle{ \frac{\xiB c\mu^2 r^3}{4R_1^7}\left(1-\frac{\Rx^4}{r^4}\right)^{1/2} }& \quad \Rx< r <\Rrad \\
  \Lpre^{\max} \approx \displaystyle{ \frac{5\pi}{2} \eps L \frac{\Rx^2}{\Rrad^2} } & \qquad r \gg \Rrad
                            \end{array}\right. \quad
\end{eqnarray}
The FRB luminosity grows as $r^3$ at $r<\Rrad$ and then approaches a constant maximum value during phase~III:
\beq
\label{eq:Lpre_max}
 L_{\rm pre}^{\max} \approx 3\times 10^{38}\,L_{41}^{4/9}\mu_{33}^{2/9}\N_{37}^{2/9}\nu_3^{4/9}\eps_{-2}^{7/9}\,{\rm erg\,s}^{-1}.
\eeq 
These results demonstrate a generic feature of monster shocks: the precursor emission is suppressed at $r<\Rrad$ by the radiative losses ahead of the shock. The precursor emission does not peak near $\Rx$ where the shock is strongest. Instead, it peaks at $\Rrad\approx 10^9$\,cm where the radiative losses induced by the precursor become marginally important.

Frequency of the precursor radio wave $\nupre=\ompre/2\pi$ is given by \Eq~(\ref{eq:ompre1}), where one can substitute the solutions for $\cd$ (\Eq~\ref{eq:cd_an}) and $\gamma$ (\Eq~\ref{eq:g_an}). In particular, during phases~II and III (which dominate the burst energy), we find
\begin{eqnarray}
\label{eq:nupre_an}
  \nupre \approx \left\{\begin{array}{lr}
           \vspace*{1mm}
  \nupre^{\rm II}(\Rrad) \left(r/\Rrad\right)^{-1/2} & \quad x_1< x <\xrad \\
  \nupre^{\rm III}(\Rrad) (r/\Rrad)^2 & \qquad x \gg \xrad
                            \end{array}\right. \quad
\end{eqnarray}
where
\begin{align}
\label{eq:nupre_Rad}
   \nupre^{\rm II}(\Rrad) & \approx 0.21\,\frac{L_{41}^{17/72}\N_{37}^{29/36}\nu_3^{11/18}\eps_{-2}^{1/12}}{\mu_{33}^{25/36}}\,{\rm GHz}, \\
   \nupre^{\rm III}(\Rrad) & \approx 0.186\,\frac{L_{41}^{1/6}\N_{37}^{5/6}\nu_3^{2/3}\eps_{-2}^{1/6}}{\mu_{33}^{2/3}}\,{\rm GHz}.
\end{align} 
Since the emitted burst energy peaks at $r\approx\Rrad$, its characteristic frequency should also be defined at $\Rrad$. Note that $\nupre(\Rrad)$ extrapolated from phases~II and III nearly match, although they are not exactly equal. 

The analytical expressions for $\Lpre$ and $\nupre$ (\Eqs~(\ref{eq:Lpre_an}) and (\ref{eq:nupre_an})) give useful approximations to the numerical results, as demonstrated in Figure~\ref{fig:sim}. These expressions can be used to scale the results to different parameters of the magnetar ($\mu$ and $\N$) and the magnetosonic perturbation ($L$ and $\nu$). 

Finally, consider the duration of the predicted FRB.
The observed time of shock expansion to a given radius $r$ is given by 
\beq
  \tobs(r)=\int_{\Rx}^r \frac{dr}{2c\gsh^2}=\int_{\Rx}^r \frac{\gamma\,dr}{c\sigbg \cd^2}.
\eeq
In particular, the duration of phase~II is
\beq
  \Delta\tobs^{\rm II}\approx \frac{16\pi \me c}{\mu^2} \sqrt{\frac{\N}{\sT\eps}} \,R_1^5 \left[1-\left(\frac{R_1}{\Rrad}\right)^{1/4}\right],
\eeq
where we used $\sigbg r^3=\sigx\Rx^3=(\mu^2/4\pi\me c^2\N)$. The last factor in square brackets approximately equals $0.15$ and weakly depends on the parameters of the explosion. This gives
\beq
  \Delta\tobs^{\rm II}\approx 0.0183 \,\frac{\mu_{33}^{1/7}\N_{37}^{1/7}}{L_{41}^{5/14}\nu_3^{5/7}\eps_{-2}^{1/7}}\,{\rm ms}. 
\eeq

The duration of phase~III is
\beq
   \Delta\tobs^{\rm III}\approx \frac{\Rrad}{4c\gsh^2(\Rrad)}
   \approx 0.0265\,\frac{\mu_{33}^{1/9}\N_{37}^{1/9}}{L_{41}^{5/18}\nu_3^{7/9}\eps_{-2}^{1/9}}\,{\rm ms}. 
\eeq

%############################################################################

\section{Damping of the emitted radio burst}
\label{damping}

The produced FRB escapes to distant observers if it successfully propagates through a broad range of radii, including the outer magnetosphere and the wind outside the light cylinder. The FRB faces greatest danger if the ambient plasma responds to it in the regime of stochastic heating \citep{Beloborodov22}, which involves fast irreversible absorption of the wave energy (see also \cite{Lyubarsky18}) and radiative losses of the heated particles. Fortunately for the radio waves emitted by monster shocks, stochastic heating may be avoided under certain conditions given below. Then, the plasma at all radii responds to the wave in the regular oscillation regime, which involves only mild damping of the wave. Below we first evaluate damping in the regular oscillation regime and then discuss the dangerous onset of stochastic heating.

\subsection{Damping in the regular oscillation regime}

Regular oscillation means that the velocity of a plasma particle in the radio wave simply tracks the variation of the wave potential. The plasma returns to its original state after each oscillation, with zero energy consumption from the wave, if the particles experience no radiative losses during the oscillation. Radiative losses create non-zero damping of the wave.

The power radiated by each oscillating particle $\dEe$ is given in \Sect~\ref{deceleration}. It is Lorentz invariant, however its interpretation depends on the reference frame. In particular, consider radii $r\sim (1-10)\Rx$, where the plasma flow on the plateau ahead of the monster shock has a huge bulk Lorentz factor $\gD$. When viewed in the static lab frame, the net radiative losses induced by the precursor far exceed the precursor wave energy and hence $\dEe$ represents the energy losses of the plasma flow, as described in \Sect~\ref{deceleration}. By contrast, when viewed in the local fluid frame $\tKF$, the precursor wave dominates (and controls) the plasma energy density, which has no bulk motion component, $\tilde\beta_{\rm D}=0$. In this frame, $\dEe$ expresses losses (per particle) of the radio wave that sustains the particle oscillations. So, the energy losses of the radio wave should be calculated in frame $\tKF$.

When expressed in frame $\tKF$, the average radiated power per particle is $\dEe=c\sT \tUpre \tu_\xi^2 = c\sT \tUpre(1+a^2)$. It may be rewritten as
\beq
   \dEe=\sigsc \tF,   \qquad  \sigsc = (1+a^2)\sT,
\eeq
where $\tF\approx c\tUpre$ is the energy flux of the wave measured in the fluid frame, and $\sigsc$ is the effective scattering cross section for the wave. Scattering here means conversion of the radio wave into the radiation produced by the oscillating particles. The scattering cross section $\sigsc$ is defined in the fluid rest frame.

Consider now a layer $(\xi,\xi+d\xi)$ of the radio precursor. It was emitted by the shock at some radius $\rem(\xi)$ with strength parameter $\aem(\xi)$. As the radio wave propagates to radii $r>\rem$, its strength parameter decreases: 
\beq
\label{eq:aem}
  a(r,\xi)=\frac{\rem(\xi)\aem(\xi)}{r}.
\eeq 
The number of plasma particles crossing the shock precursor per unit area of the shock per unit time in the lab frame is $c\nbg$, and the scattering optical depth encountered by the precursor layer $\xi$ is\footnote{Scattering rate $c(1-\bD)\sigsc n$ gives optical depth $d\tausc=\sigsc \nbg dr$ when the wave propagates distance $dr$, since $n=\nbg/(1-\bD)$ for any quasi steady profile of plasma density ahead of the shock.}
\beq
   \tausc(\xi)=\int_{\rem}^r \sigsc \nbg\,dr \approx \frac{\sT \N \aem^2}{4\rem^2},
\eeq
where we substituted $\nbg=\N r^{-3}$ and used $a\gg 1$. Note that the contribution to $\tausc$ from propagation to large $r\gg\rem$ is small; half of $\tausc$ is encountered as the precursor wave propagates from $\rem$ to $2^{1/4}\rem$.

The strength parameter of the emitted precursor (\Eq~\ref{eq:a}) may be expressed as
\beq
\label{eq:a1}
   \aem=\frac{1}{3}\,\xiB^{1/2}\, \gamma(\rem) \, \cd(\rem),
\eeq
where we used the transformation of the upstream Lorentz factor to the downstream $\gud=\gamma\cd$. In particular, for the precursor layers emitted during phase~II, we find
\beq
\label{eq:a2}
   \aem\approx \frac{2 \rem}{3\sqrt{\sT \N}} \;\;\; (R_1<\rem<\Rrad),
\eeq
which gives $\tausc\approx 1/9$. For phase~III, using the analytical solutions for $\gamma$ and $\cd$, we find $\tausc\approx (2/9)(\Rrad/\rem)^8$. Thus,
\beq
   \tausc\approx\left\{\begin{array}{lr}
   \vspace*{2mm}
    \displaystyle{ \frac{1}{9} } & R_1<\rem<\Rrad \\
     \displaystyle{ \frac{2}{9} \frac{\Rrad^8}{\rem^8} } & \rem\gg\Rrad
                                                           \end{array}\right.
\eeq
The small $\tausc\ll 1$ shows that the radio waves produced by monster shocks are weakly damped, as long as their interaction with the upstream plasma occurs in the regular oscillation regime. The above calculation also assumed that the magnetospheric plasma particles cross the precursor of thickness $c(\xish-\xi_0)\approx \pi/\omega$ on a timescale $\tcross<r/c$.

\subsection{Stochastic heating}
\label{stochastic}

Plasma response to the radio wave changes from regular oscillations to stochastic heating when the parameter $b\equiv\tomB/\tompre$ exceeds $\bcr\approx (1/3)\sqrt{1+a^2}$ (see Figure~3 in \citetalias{Beloborodov26}). In the relevant regime of $a\gg 1$, this condition becomes 
\beq 
\label{eq:stochastic}
   \frac{b}{a}=\frac{\tBbg}{\tErms}>\frac{1}{3} \quad ({\rm stochastic~heating};\, a\gg 1).
\eeq
where $\tErms\equiv\langle \delta\tilde{E}^2\rangle^{1/2}$. When this condition becomes satisfied, the particles begin to absorb energy from the wave and develop gyration  in the fluid frame with a growing Lorentz factor $\tg\gg a$. Thus, stochastic heating can be thought of as synchrotron absorption, which includes a systematic gain (\cite{Lyubarsky18}, see also \cite{Sobacchi25}) and a random walk in $\tg$ \citep{Beloborodov22}, broadening the particle distribution function, see Appendix~D in \citetalias{Beloborodov26}. 
This results in a fast growth of plasma energy per particle $\gstoch\me c^2$.

For a plasma exposed to a radio wave with frequency $\nupre$ and cumulative energy distribution $\E(\xi)$, one finds \citepalias{Beloborodov26}:
\beq
\label{eq:gstoch}
  \gstoch\approx \left(\frac{r_e\EFRB}{\pi \me c\,\nupre r^2}\right)^{3/7} \!\! b^{1/7}
  \! \approx\! \frac{1.7 \! \times \! 10^4 \,\E_{\rm FRB,33}^{3/7}b_2^{1/7}}{(\nupre/{\rm GHz}) \,r_9^{6/7}},
\eeq
where $r_e=e^2/\me c^2$. The large $\gstoch\gtrsim 10^4$ leads to radiative cooling of the plasma, which can balance stochastic heating at $\tg=\grad\sim 10^4$ \citepalias{Beloborodov26}, continuing to drain the FRB energy. Unlike regular oscillations, stochastic heating (and the accompanying radiative losses) is irreversible. The isotropic equivalent of energy spent by the FRB on stochastic heating may be estimated as 
\beq 
   \Delta\E\sim 4\pi \N \c \gstoch\me c^2.
\eeq
Here $4\pi\N$ is the number of magnetospheric particles exposed to the FRB (isotropic equivalent), and the factor $\c=\gD(1+\bD)$ transforms the consumed energy from the fluid frame to the lab frame. Fluid motion in the radio precursor ahead of the monster shock has $\c\approx a/2\gamma$, which steeply grows with $r$.  At radii $r\gtrsim 10^9$\,cm (after most of the FRB energy $\EFRB$ has been generated), the above expression for $\Delta\E$ exceeds $\EFRB$, which means strong damping of the FRB.

In view of this damping threat one needs to check whether the FRB produced by the monster shock actually encounters the stochastic heating regime $b/a>1/3$. This is done in Appendix~\ref{app:stochastic}. For this analysis we relax the assumption used in previous sections: we no longer assume that the density parameter $\N=r^3\nbg$ is a global constant and allow $\N$ to depend on $r$. This extension of the model is important when considering FRB propagation over  a broad a range of radii, significantly exceeding the FRB production radius $r\sim \Rrad$.

The results derived in Appendix~\ref{app:stochastic} may be summarized as follows. The bottleneck for escape is radius $r\sim \RLC$, and the outcome is determined by the parameter $\NLC\equiv\N(\RLC)$. The FRB escapes if $\NLC$ is significantly lower than the fiducial $\N=10^{37}$ that we assumed in the emission region $r\sim\Rrad$. In particular, for Model~W we find that stochastic heating does not occur at any $r\lesssim\RLC$ and the FRB escapes the magnetosphere if 
\beq
    \NLC\lesssim 4\times 10^{35} \qquad {\rm (escape~in~Model~W).}
\eeq
As shown in Appendix~\ref{app:stochastic}, this condition also allows the FRB to escape through the wind that flows outside the magnetosphere at $r>\RLC$.

%#########################################################################

\section{Discussion}
\label{discussion}

This paper shows that magnetospheric shocks in magnetars naturally produce FRBs in the GHz band with energies $\EFRB\sim 10^{34}$\,erg. Although the shock is strongest near radius $\Rx\sim 10^8$\,cm, the FRB production peaks at $\Rrad\approx 10^9\,{\rm cm}\gg\Rx$, when the shock weakens by a few orders of magnitude. This is caused by of a self-regulation process that involves interaction of the shock precursor with the upstream plasma. 

The self-regulation also results in the weak dependence of $\EFRB$ on the energy of the primary kHz disturbance $\E$ that launched the shock. We find that $\EFRB$ scales with $\E$ approximately as $\E^{0.2}$, and the efficiency of FRB emission is 
\beq
  \eta\equiv\frac{\EFRB}{\E}\approx 10^{-4}\,\E_{38}^{-0.8}.
\eeq 
The reduction of $\eta$ at large $\E$ is caused by large radiative losses of the upstream plasma flow exposed to the radio precursor before reaching the shock. The low $\eta$ at large $\E$ implies that monster shocks inside the magnetosphere cannot  explain bright cosmological FRBs with $\EFRB\gtrsim 10^{38}$\,erg. Shocks become bright FRB sources only when they expand into the wind far beyond the magnetosphere \citep{Beloborodov20}. 

The magnetospheric shocks are natural sources of weak FRBs similar to those observed in SGR~1935+2154. Several predictions of the calculated model with a kHz magnetospheric disturbance with $\E\sim 10^{38}$\,erg appear consistent with radio  and X-ray observations of SGR~1935+2154 \citep{CHIME20,Mereghetti20}:
\medskip
\\
(1) FRB frequency $\sim 1$\,GHz.
\medskip
\\
(2) FRB energy $\sim 10^{34}$\,erg.
\medskip
\\
(3) Sub-ms duration of the radio spikes.
\medskip
\\
(4) Energy of the X-ray spikes accompanying the FRBs, $\E_{\rm X}\sim 10^{38}$\,erg. The observed $\E_{\rm X}$ is consistent with the prediction that  monster shocks radiate in hard X-rays about half of the kHz disturbance energy $\E$ \citepalias{Beloborodov23}.  
\medskip
\\
(5) The $\sim 5$\,ms delay of the X-ray spike relative to the FRB is consistent with the expected delay $\sim \Rx/c$. The X-rays trail the GHz precursor because they are produced behind the shock 
\citepalias{Beloborodov23}. 
\medskip
\\
(6) FRBs rarely accompany X-ray bursting activity. Radio waves produced by magnetospheric shocks usually get absorbed on the way out, before reaching $\RLC$. The FRB escapes if the plasma density where the line of sight crosses the light cylinder happens to be smaller by a factor of $\sim 30$ compared to a fiducial value expected for active magnetars.
\medskip
\\
The duration $\delta t$ of an individual radio burst predicted by the model is shorter than the observed durations. The calculated example model gives $\delta t\lesssim 0.3$\,ms, with the main peak produced within 0.1\,ms. The predicted $\delta t$, however, depends on the temporal structure of the kHz magnetospheric disturbance, which can launch multiple shocks and produce an FRB with multiple peaks with an overall duration that significantly exceeds 1\,ms. In addition, the observed $\delta t$ might be affected by propagation effects at large distances from the magnetar.

Calculations in this paper have some limitations. In particular, we focused on FRB generation by shocks near the equatorial plane of a dipole magnetosphere. This choice was made for two reasons: (i) calculations simplify in the equatorial plane, and (ii) global kinetic simulations of monster shocks \citep{Bernardi25} suggest that the precursor emission occurs preferentially near the equatorial plane and becomes suppressed closer to the polar axis. The assumption of a dipole background magnetosphere at radii $r>10^8$\,cm is not unreasonable but may be relaxed in future models; it is possible that the outer magnetosphere is significantly deformed from the dipole configuration by a global twist. Furthermore, it can contain significant turbulent motions excited by small quakes preceding the strong kHz perturbation. The pre-explosion turbulence can significantly corrugate the monster shock \citep{Grehan26} and change its radiative properties.

More work may be needed to study the microphysics of monster shocks and their precursor emission using kinetic plasma simulations \citep{Vanthieghem25a,Bernardi25}. In particular, simulations of extremely strong shocks with radiative losses could reveal new features not seen earlier.  

\medskip

The author acknowledges support by NASA grants 21-ATP21-0056 and 80NSSC24K1229, NSF
grant AST-2408199, and Simons Foundation grant 446228. This work was facilitated by Multimessenger Plasma Physics Center (MPPC) grant PHY-2206609.

%#############################################################################

\appendix

\section{Analytical solution}
\label{derivation}

\subsection{Downstream compression factor $\cd$}

A key evolving parameter of the shock is its downstream velocity $\betad$ or the corresponding compression factor $\cd=\gd(1+\betad)$. It controls the precursor luminosity (\Eq~\ref{eq:Lpre}) and also enters the expression for the precursor frequency (\Eq~\ref{eq:ompre1}). The parameter $\cd$ is a known function of $x=r/\Rx$ and $\xish=\tobs$ (\Eq~\ref{eq:cd}), so its evolution with radius $r\approx ct$ is determined by the shock trajectory $\xish(t)$. 

Simple analytical expressions for $\xish(t)$ and shock Lorentz factor $\gsh(t)$ hold in the region where $\cd\ll 1$ (\Eqs~(\ref{eq:xish_appr}) and (\ref{eq:gsh_appr})). Then, we find $\cd\approx \gsh/\gshd$ from the jump condition~(\ref{eq:jump}):
\beq
\label{eq:cd_appr}
  \cd^2\approx \frac{2\gamma\gsh^2}{(1+\chi)^{2/7}\sigbg}
  \approx \frac{\gamma}{(1+\chi)^{2/7}} \frac{\omega r }{2c\sigbg} \sqrt{x^4-1},
\eeq 
where $x\equiv r/\Rx$.
The Lorentz factor of particles after crossing the radiative shock is $\gamma_e=\gamma/(1+\chi)^{2/7}$ (\Sect~\ref{losses_shock}). With no radiative losses (in the precursor or the shock itself), the particles would have $\gamma_e=\gamma_0=\Wp\sigbg/r$, and the condition $\cd\ll 1$  may be stated as  
\beq
\label{eq:cd1}
   \cd^2\approx  \frac{\gamma_e}{2\gamma_0} \frac{\omega\Wp}{c} \sqrt{x^4-1}\ll 1.
\eeq
One can see that shocks with strong radiative losses, $\gamma_e\ll\gamma_0$, have $\cd\ll 1$ in a region of $x\simgt 1$.

A different expression for $\cd$ holds during a later evolution phase when the shock shifts to the part of the magnetosonic pulse with $E>0$ ($\omega\xish>2\pi$). Here, $\betad=\Ed/(\Bbg+\Ed)$ approaches unity, since $\Ed/E_0$ grows and $\Bbg\propto r^{-3}$  becomes much smaller than $\Ed$. As a result, the shock position $\xish$ stalls at some $\xi_\star$ slightly larger than $2\pi/\omega$ (Figure~\ref{fig:xish}). In this late phase, $\cd\gg 1$ and the evolution of $\cd$ with radius is given by
\beq
\label{eq:cd2}
   \cd\approx x \, \sqrt{\sin(\omega\xi_\star)} \qquad ({\rm when}~\cd\gg 1).
\eeq

\subsection{The coupled evolution of $\cd$ and $\gamma$}

The coupled evolution of the shock and its precursor is essentially the coupled evolution of $\gamma$ and $\cd$: $\cd$ controls the precursor luminosity, which in turn controls $\gamma$. After the shock forms at radius $\Rx$, the precursor almost immediately becomes sufficiently strong to reduce $\gamma\ll\gamma_0$. This effect is described by \Eq~(\ref{eq:dyn}), which has three terms. Three distinct phases of shock+precursor evolution are defined by the relative importance of these three terms.

{\bf Phase I:} $1<x<x_1$. The upstream Lorentz factor $\gamma$ is regulated so that the precursor deceleration effect balances the plateau acceleration, i.e. the two terms on the right-hand side of \Eq~(\ref{eq:dyn}) nearly balance each other (and far exceed the term on the left-hand side):
\beq
\label{eq:eq}
   \frac{2\sT\gamma^2(\xi)\Upre(\xi)}{\me c} \approx \frac{c\sigbg}{r}.
\eeq
This balance holds across the precursor $0<\xi<\xish$. At $\xi=\xish$ it sets the upstream Lorentz factor at the shock,
\beq
\label{eq:gam_I}
  \gamma\approx \frac{r}{\cd^2}\sqrt{\frac{2}{\sT\N\xiB}}     \qquad (1<x<x_1),
\eeq
where we used \Eq~(\ref{eq:Upre}) for $U(\xish)$. During this evolution phase $\cd\ll 1$, so one can use \Eq~(\ref{eq:cd_appr}). Substituting \Eq~(\ref{eq:gam_I}) into \Eq~(\ref{eq:cd_appr}), we find
\beq
\label{eq:cd_I}
  \cd^4 \approx \sqrt{\frac{\N}{2\sT\xiB}}  \frac{4\pi \me c\, r^7}{\mu^2\Rx^2 }\frac{\sqrt{1-x^{-4}}}{(1+\chi)^{2/7}}.
\eeq
This phase continues as long as $d\gamma/d\xi$ is small compared to both terms on the right-hand side of \Eq~(\ref{eq:energy1}). It ends at $x_1=R_1/\Rx$ determined below. 

\medskip

{\bf Phase II}: $x_1<x<\xrad$. Here, the deceleration term in \Eq~(\ref{eq:dyn}) becomes dominant, and the acceleration term $c\sigbg/r$ may be neglected. Then, \Eq~(\ref{eq:dyn}) can be immediately integrated, and one finds that the upstream particles reach the shock with Lorentz factor,
\beq
\label{eq:dec1}
    \gamma\approx \frac{\me c^2}{2\sT \Sigma}, \qquad \Sigma\equiv \int_{\xi_0}^{\xish} \Upre c\, d\xi
    \qquad (\gamma\ll\gu).
\eeq

With approximately radial expansion of the precursor, its column energy density $\Sigma$ experiences dilution $\propto r^{-2}$. The change of $\Sigma$ during time $dt$ is described by 
\beq
\label{eq:dSigma}
   d(r^2\Sigma)=r^2 \dot{\Sigma}\, dt, 
\eeq
where $\dot{\Sigma}$ is the rate of precursor injection by the shock. During time $dt$ the shock adds a new precursor layer $d\xi\approx dt/2\gsh^2\approx dr/2\gsh^2c$ with density $U$ given by \Eq~(\ref{eq:Upre}). So, $\dot{\Sigma}\,dt=U c\, d\xi = U dr/2\gsh^2$, and \Eq~(\ref{eq:dSigma}) gives the following evolution equation for $\Sigma$:
\beq
   \frac{2\gsh^2}{r^2}\,\frac{d}{dr}\left(r^2\Sigma\right)=U
   =\frac{\xiB \Bbg^2\cd^4}{16\pi}.
\eeq
Substituting $\Sigma\approx\me c^2/2\sT\gamma$ (from \Eq~(\ref{eq:dec1})), we obtain the evolution equation for the Lorentz factor $\gamma(r)$ of particles that enter the shock (after crossing its precursor) at radius $r$ or time $t\approx r/c$:
\beq
\label{eq:dec}
   \frac{1}{r^2}\, \frac{d}{dr} \left(\frac{r^2}{\gamma}\right)
    \approx \frac{\sT\Bbg^2 \xiB\cd^4}{16\pi\me c^2 \gsh^2}.
\eeq 
The shock Lorentz factor satisfies $\gsh=\gshd\cd$ (\Eq~\ref{eq:xish_diff1}) where $\gshd^2\approx \sigbg/2\sigma$ is set by the jump condition (with $\chi\ll 1$, strongly reduced at this phase). This gives
\beq
\label{eq:gsh2}
   \gsh^2 \approx \frac{\sigbg\cd^2}{2\gamma},
\eeq
and \Eq~(\ref{eq:dec}) becomes 
\beq
\label{eq:dec2}
   \frac{1}{r^2}\, \frac{d}{dr} \left(\frac{r^2}{\gamma}\right)
    \approx \frac{\sT\Bbg^2 \xiB \gamma\cd^2}{8\pi\me c^2 \sigbg}
    =\frac{1}{2}\,\sT\xiB \nbg \gamma\cd^2.
\eeq 
It can be integrated for $\gamma/r^2$:
\beq
  \frac{r^4}{\gamma^2} \approx \sT \xiB\,\N \int_{\Rx}^{r}  \cd^2\, r \,dr.
\eeq
Since $\cd^2$ grows with $r$, the integral peaks at the upper limit, 
\beq
    \int_{\Rx}^{r}  \cd^2\,r \,dr \approx \frac{r^2\cd^2}{4},
\eeq
and we obtain 
\beq
\label{eq:gam_II}
  \gamma^2 \approx \frac{4 r^2}{\sT \N \xiB\,\cd^2} \qquad (x_1<x<\xrad).
\eeq

One can now check when phase~II begins, i.e. when the deceleration term in \Eq~(\ref{eq:dyn}) becomes dominant over the acceleration term $c\sigbg/r$. The ratio of the two terms at the shock is
\beq
    \frac{2\sT \gamma^2 r U}{\me c^2 \sigbg} 
    = \frac{\sT \xiB\Bbg^2 r \gamma^2\cd^4}{8\pi\me c^2 \sigbg} 
    = \frac{\sT \N \xiB}{2r^2}\, \gamma^2\cd^4.
\eeq
Using \Eq~(\ref{eq:gam_II}), we find that phase~II occurs when
\beq
    \frac{2\sT \gamma^2 r U}{\me c^2 \sigbg}
    \approx  2\cd^2\gg1.
\eeq
The transition between phases~I and II is near radius $x_1=R_1/\Rx$ where $\cd=1$. 

Remarkably, the simple extrapolation of the expression for $\cd\ll 1$ in phase~I (\Eq~\ref{eq:cd_I}) to phase~II holds with good accuracy (Figure~\ref{fig:sim}). It can be simplified using $(1+\chi)^{2/7}\approx 1$ and $x^4-1\approx x^4$ during phase~II. Using this equation, we find the radius $x_1$ where $\cd=1$ 
\beq
   x_1^7\equiv\left(\frac{R_1}{\Rx}\right)^7\approx  \sqrt{\frac{2\sT\xiB}{\N}} \frac{\mu^2}{4\pi \me c\,\omega \Rx^5  }.
\eeq
This gives $x_1=4.41$ and $3.02$ for Models~S for W, respectively.

The approximate solution for $\gamma$ (\Eq~\ref{eq:gam_II})  holds at $x>x_1$ as long as radiative losses are strong, i.e. $\gamma\ll\gamma_0$ where $\gamma_0\approx (\pi c\sigbg/r\omega)$ at $x\gg 1$ (\Eq~\ref{eq:gamma0}). The end point of phase~II, $\xrad=\Rrad/\Rx$, may be estimated from the condition $\gamma\sim\gamma_0$:
\beq
   \gamma^2\approx \frac{4r^2}{\sT \N \xiB\,\cd^2} 
   \approx \frac{\pi^2 c^2\sigbg^2}{r^2\omega^2}\approx\gamma_0^2,
\eeq
which gives 
\beq
   \xrad^8=\frac{\sT\xiB L^2}{2\me^2 c^4\N\omega^2}\,\frac{\cd^2}{r^2}.
\eeq
At this point, $\cd^2$ significantly exceeds unity, $\cd^2/r^2\approx \sin(\omega\xish)/\Rx^2$ (\Eq~\ref{eq:cd}). So, we obtain
\beq
    \xrad^8\equiv\left(\frac{\Rrad}{\Rx}\right)^8=\frac{\sqrt{2} \sT\xiB L^{5/2}\sin(\omega\xish)}{\me^2 c^{9/2}\N\omega^2\mu}.
\eeq
Note that $\sin(\omega\xi)\sim 0.1$ at $x=\xrad$ (Figure~\ref{fig:xish}); an analytical approximation for $\sin(\omega\xish)$ is derived below. Phase~II exists if $\xrad>x_1$. In particular, we find $\xrad\approx 10.3$ for Model~S and $\xrad\approx 5.7$ for Model~W. 

\medskip

{\bf Phase III}: $x\gg \xrad$. Here, the upstream Lorentz factor is weakly affected by radiative losses, and so 
\beq
   \gamma\approx\gamma_0\approx  \frac{\pi c\sigbg}{\omega r}.
\eeq
The large value of $\cd$ in phase~III allows one to simplify \Eq~(\ref{eq:cd}) to 
\beq
   \cd^2\approx x^2\sin(\omega\xish)\gg 1.
\eeq
The corresponding shock Lorentz factor (\Eq~\ref{eq:gsh2}) is 
\beq
\label{eq:gshIII_}
  \gsh^2\approx\frac{\sigbg\cd^2}{2\gamma_0}
  \approx \frac{\nu \Rx}{c} \sin(\omega\xish) \,x^3 \quad\; (x \gg \xrad).
\eeq
Then \Eq~(\ref{eq:xish_diff1}) describing shock propagation takes the form
\beq
    \frac{d\xish}{dx}\approx \frac{\pi}{\omega \sin(\omega\xish) \,x^3}.
\eeq
One can integrate it for $\xish(x)$ and find
\beq
      \psi^2\approx \psi_\star^2 - \frac{\pi}{x^2} \qquad (\psi\equiv \omega\xish-2\pi\ll 1).
\eeq
The onset of this solution occurs at $x_0^2\approx 2\xrad^2$ (Figure~\ref{fig:xish}). At larger $x>x_0$, $\psi(x)$ quickly asymptotes to $\psi_\star$, which is not much greater than $\psi_0\equiv \psi(x_0)$: we find $\psi_0^2/\psi_\star^2\approx 0.6$ both Models~S and W. This gives 
\beq
  \sin(\omega\xi_\star) \approx \frac{\sqrt{5\pi}}{2\xrad},
 \qquad \cd^2\approx \frac{\sqrt{5\pi}}{2\xrad}\, x^2.
\eeq

\subsection{The quasi steady approximation}
\label{app:steady}

\Eq~(\ref{eq:dyn}) for the dynamics of plasma exposed to the radio precursor is equivalent to \Eq~(51) in \citetalias{Beloborodov26}. It describes a quasi steady pattern of plasma flow in a radio wave packet --- the ``compression front.'' The equation holds in the limit of $t_\star\ll r/c$, where $t_\star$ is the timescale for relaxation toward the steady pattern. Below we discuss this approximation.

The relaxation process may be viewed as propagation of MHD information across the radio wave packet \citepalias{Beloborodov26}. It involves magneto-sound (with wavelength comparable to the width of the packet), which propagates relative to the fluid with Lorentz factor
\beq
   \gs\approx \sqrt{1+\sigma_{\rm eff}}\gg 1,
\eeq
where
\beq 
  \sigma_{\rm eff}=\frac{\tB^2}{4\pi\tg\trho c^2} = \frac{\c\sigbg}{\sqrt{1+a^2}}.
\eeq
Magneto-sound can propagate in both directions (with speeds $\pm\bs$), and one can define $\c_{\rm s}^\pm\equiv\gs(1\pm\bs)\approx(2\gs)^{\pm 1}$. The corresponding speed in the lab frame is described by $\c_\pm=\c\c_{\rm s}^\pm$. Since $\c_+\gg\c_-$, the spreading of MHD information relative to the radio wave packet is slowest in the + direction. It occurs with relative speed 
\beq
  \frac{d\xi_+}{dt}=1-\beta_+
   =\frac{2}{1+\c_+^2}
  \approx \frac{2}{1+4\gs^2\c^2}. 
\eeq
Thus, the timescale for MHD communication across the packet may be estimated as 
\beq
\label{eq:t_+}
   t_+=\int\frac{d\xi}{1-\beta_+}
   \approx \int \left( \frac{1}{2}+\frac{2\c^3\sigbg}{\sqrt{1+a^2}}\right)d\xi. 
\eeq 
For an MHD fluid with small radiative losses, $\beta_\pm$ have the rigorous meaning of the speeds of $\C^\pm$ characteristics \citep{Beloborodov24}. When radiative losses are significant, the MHD equations cannot be cast into the standard form of Riemann invariants along $\C^\pm$; however, one can still think of $\beta_\pm$ as characteristic speeds of MHD information propagation, and \Eq~(\ref{eq:t_+}) gives a suitable estimate for the relaxation time $t_\star$ as seen in numerical simulations \citepalias{Beloborodov26}.

In the case of FRB interaction with a magnetar wind considered in \citetalias{Beloborodov26}, the steady state approximation is easily violated, $ct_+/r>1$. By contrast, in the case of a monster shock precursor, the quasi steady state holds, $ct_+/r<1$. This can be verified using \Eq~(\ref{eq:t_+}) with the obtained numerical solutions for $\c\approx a/2\gamma$ and $a(\xi)$. One can also estimate the integral in  \Eq~(\ref{eq:t_+}) analytically as follows:
\beq
  \frac{ct_+}{r}\sim \left( \frac{1}{2}+\frac{\c^2\sigbg}{\gamma}\right)\frac{c\,\Delta\xi}{r}
  \sim \frac{c}{\omega r} + \frac{\c^2\gu}{\gamma}.
\eeq
Both terms on the right hand side are smaller than unity. In particular, at the shock one can use $\gDud\approx 3\xiB^{-1/2}\approx 30$ (\Eq~\ref{eq:gDud}), which implies $\c\approx \cd/\gDud\approx \cd/30$. This gives $\c^2\gu/\gamma\ll 1$ in both models shown in Figure~\ref{fig:sim}.

%#############################################################################

\section{Onset of stochastic heating}
\label{app:stochastic}

The FRB emitted by the monster shock will be absorbed on its way out of the magnetosphere if the ambient plasma responds to the radio wave in the stochastic heating regime (section~\ref{stochastic}). This regime occurs when $b/a>1/3$ (\Eq~\ref{eq:stochastic}). Below we investigate if this condition is met at some radius. We first consider FRB propagation in the closed magnetosphere at $r<\RLC$ and then in the wind zone $r>\RLC$.

\subsection{FRB propagation at $r<\RLC$}
\label{stoch}

We continue to view the problem in coordinates $(t,\xi)$ or, equivalently, $(r,\xi)$. Here $r\approx ct$ is the radius of the thin shell containing the kHz magnetosonic perturbation, the shock, and the produced FRB, all propagating outward with nearly speed of light. The coordinate $\xi\ll r/c$ is used to resolve the structure of this shell. The FRB occupies the region $\xi_0<\xi<\xish$ (Figure~\ref{fig:plateau}). 

Consider a snapshot of this propagating shell when it has reached a radius $r$.
First, let us evaluate $b/a$ in the FRB layer that is currently being emitted by the shock, i.e. the precursor layer at $\xi=\xish$. When viewed from the downstream frame $\KFd$, the precursor electric field satisfies $\dErms/\Bud=\xiB^{1/2}$ (\Eq~\ref{eq:Epre}) where $\dErms\equiv \langle \Epred^2\rangle^{1/2}$. The corresponding ratio $\tErms/\tB$ measured in the upstream fluid frame $\tKF$ is determined by field transformations:  $\tB\approx 2\cud\Bud$ (\Eq~\ref{eq:Bd}) and
 $\tEpre=\Epred/\cud$ (\Eq~\ref{eq:prec_transf}), 
 where $\cud=\c/\cd$ describes the relative motion of the two frames (\Sect~\ref{frames}). 
This gives
\beq
  \left(\frac{\tErms}{\tB}\right)_{\rm em} \approx \frac{\xiB^{1/2}}{2\cud^2}, \qquad \cud\approx \frac{\xiB^{1/2}}{6}.
\eeq
The second equality stated in this equation follows from \Eq~(\ref{eq:gDud}) and $\cud\approx(2\gDud)^{-1}$. Subscript ``em'' indicates that here we are evaluating the precursor wave at its emission location, right at the shock. Then, we find 
\beq
\label{eq:b_a_em}
  \left(\frac{b}{a}\right)_{\rm em}\approx \frac{\xiB^{1/2}}{18}.
\eeq
This ratio is far below 1/3, so the emitted precursor begins to propagate through the upstream plasma in the regular oscillation regime.

Next, consider the emitted precursor layer $\xi$ as it continues to propagate to radii $r$ significantly larger than its emission radius $\rem(\xi)$ (while the shock is shifting to larger $\xish$). 
We wish to know if the layer $\xi$ encounters the condition $b/a>1/3$ at some $r>\rem$.
It will be helpful to express $b/a=\tErms/\tB$ in terms of the fields measured in the lab frame. Transformations $\tErms=\Erms/\c$ and $\tB=\c\Bbg$ [where $\c(r,\xi)=\gD(1+\bD)$ describes the motion of the local fluid frame] give
\beq
  \frac{b}{a}(r,\xi)=\c^2(r,\xi)\frac{\Bbg}{\Erms}\approx \frac{a^2}{4\gamma^2}\frac{\Bbg}{\Erms}.
\eeq
One can see that $b/a$ in the layer $\xi$ changes with $r$ from its initial value at $\rem$ because (i) $\gamma(r,\xi)$ changes from its value $\gem$ at $\rem$, (ii) the strength parameter of the radio wave changes from $\aem$ as $a(r,\xi)=\aem \rem/r$, and  (iii) the ratio $\Bbg/\Erms$ decreases as $r^{-2}$. The resulting $b/a$ can be expressed as
\beq
\label{eq:b_a1}
   \frac{b}{a}(r,\xi)\approx \left(\frac{b}{a}\right)_{\rm em} \! S^2,
   \qquad S(r,\xi) \equiv  \frac{\rem^2}{r^2}\frac{\gem}{\gamma(r,\xi)}.
\eeq
The Lorentz factor of particles at the shock, $\gem$, is given by \Eq~(\ref{eq:g_an}), and the solution for $\gamma(r,\xi)$ anywhere in the precursor wave packet is described in Appendix~\ref{derivation}. This allows one to find $b/a$ at each $\xi$ as the shock and its precursor propagate to larger $r$. Let us examine three phases of this propagation.   

During phase~I ($\Rx<r<R_1$), $\gamma(r,\xi)$ satisfies \Eq~(\ref{eq:eq}):
\beq
  \gamma^2 \approx \frac{\me c^2\sigbg}{2\sT r \Upre (r,\xi)}
  \approx \gem^2\frac{\rem^4}{r^4}\frac{\Upre_{\rm em}}{\Upre}
   = \gem^2\frac{\rem^2}{r^2}.
\eeq
It gives $b/a=(b/a)_{\rm em} \rem^2/r^2<(b/a)_{\rm em}\ll 1/3$, so stochastic heating by the precursor cannot occur during phase~I.

During phase~II ($R_1<r<\Rrad$), $\gamma(r,\xi)$ is given by 
\beq
\label{eq:g_II}
    \frac{1}{\gamma}\approx \frac{1}{\gamma_0} 
    +\frac{2\sT \Sigma}{\me c^2}, \quad \Sigma(r,\xi) \equiv \int_{\xi_0}^{\xi} \Upre(r,\xi') c\,d\xi',
\eeq
where 
\beq
\label{eq:g0_}
   \gamma_0(r,\xi)\approx \frac{c(\xi-\xiin)\sigbg}{r} = \frac{\xi-\xiin}{\xi_0-\xiin} \gu,
\eeq
which varies across the precursor by a factor of 2, from $\gamma_0=\gu(r)$ at $\xi=\xi_0$ to $\gamma_0\approx 2\gu(r)$ at $\xi=\xish$. The solution for $\gamma^{-1}$ (\Eq~\ref{eq:g_II}) implies
\beq
\label{eq:factor}
   S \approx  \frac{\rem^2}{r^2} \frac{\gem}{\gamma_0} 
    +\frac{2\sT \Sigem\gem}{\me c^2}\frac{\rem^4}{r^4},
\eeq
where we used $\Sigma(r,\xi)=\Sigem \rem^2/r^2$ with $\Sigem(\xi)$ being the column energy density of the  precursor wave packet at radius $\rem(\xi)$. At any radius $\rem>\Rx$, $\Sigem$ ahead of the shock is given by
\beq
   \Sigem(\xi) \! = \! \! \int_{\xi_0}^{\xi} \! \Upre(\rem,\xi)c\,d\xi
    = \! \! \int_{\Rx}^{\rem} \!\! \Upre'\frac{r'^2}{\rem^2} \frac{dr'}{2\gsh'^2},
\eeq
where $U'\equiv U(r',\xi)=\xiB\Bbg'^2\cd'^4/16\pi$ is the precursor energy density emitted at $r'$ (\Eq~\ref{eq:Upre}), $2\gsh'^2=\sigbg'\cd'^2/\gamma'$ (\Eq~\ref{eq:gsh_jump}), and all quantities with primes are evaluated at $r'$. Then, using $\Bbg'^2 r'^3/\sigbg'= 4\pi \N \me c^2$ we obtain
\beq
   \Sigem = \frac{\me c^2\N\xiB}{4\rem^2} \int_{\Rx}^{\rem} \gamma'\cd'^2\, \frac{dr'}{r'}.
\eeq
It is straightforward to evaluate this integral using the solutions for $\cd(r)$ and $\gamma(r)$ at the shock (\Eqs~(\ref{eq:cd_an}) and (\ref{eq:g_an})) and find
\begin{eqnarray}
\label{eq:Sigem}
  \frac{\sT \Sigem \! \gem \!}{\me c^2} \approx \! 
  \left\{\begin{array}{lr}
   \! \displaystyle{ \frac{R_1^{7/2}}{2\rem^{7/2}} \! \left[1-\frac{\Rx}{\rem}\right] \, \left[1-\frac{\Rx^4}{\rem^4}\right]^{-1/4} } \! \!
        \vspace*{2mm}
    &  {\rm (I)}\;\;\;\, \\
 \! \displaystyle{ \frac{4}{11} \! + \! \frac{R_1^{7/4}}{\rem^{7/4}} \! \left[\frac{R_1-\Rx}{\sqrt{2}\rem} \! -\! \frac{4R_1}{11\rem} \right] }  
           \vspace*{1mm}
&  \,{\rm (II)\;\;} 
                            \end{array}\right.\;\;\;\;
\end{eqnarray}
where the upper line corresponds to $\Rx<\rem < R_1$ and the lower line to $R_1< \rem <\Rrad$. Using these expressions in \Eq~(\ref{eq:factor}), one can verify that $S\lesssim 1$ for the considered interval $R_1<r<\Rrad$, and hence $b/a\ll 1/3$. We conclude that stochastic heating by the precursor does not occur during phase~II. 

Consider now phase~III ($r>\Rrad$). Ratio $b/a$ is still described by \Eqs~(\ref{eq:b_a1}) and (\ref{eq:factor}); however, now the last term $\propto\Sigem$ in \Eq~(\ref{eq:factor}) becomes negligible, as radiative losses are weak at $r>\Rrad$, so $\gamma(r,\xi)\approx\gamma_0(r,\xi)$. Using \Eq~(\ref{eq:g_an}) for $\gem=\gamma(\rem)$ and \Eq~(\ref{eq:g0_}) for $\gamma_0$, we find

\begin{eqnarray}
\label{eq:factor0}
   S \! \approx \! \frac{4\pi \me c \N r^2}{(\xi-\xiin) \mu^2} \! \left\{\begin{array}{lr}
    \displaystyle{ \sqrt{\frac{2}{\sT\Nem \xiB}}  \frac{R_1^{7/2}}{\rem^{1/2}} \!\! \left[1-\frac{\Rx^4}{\rem^4}\right]^{-1/4}  }  \!\!\!\!
        \vspace*{2mm}
     &  \rm{(I)}   \\
  \displaystyle{ \frac{2R_1^{7/4} \rem^{5/4}}{\sqrt{\sT \Nem \xiB}} }  
           \vspace*{1mm}
   &     \rm{(II)} \\
    \displaystyle{ \frac{\mu^2}{4\me c \Nem \omega \, \rem^2} }
    &   \!\!\!\!\!\! \rm{(III)}
                            \end{array}\right.  \quad\;\;\;\;
\end{eqnarray}
Here, we relaxed the approximation used in most of this paper: we no longer assume that the density parameter $\N=r^3\nbg$ is a global constant, so we distinguish between $\N(r)$ and $\Nem=\N(\rem)$. Note that at radii $r>\Rrad$ the precursor contains layers $\xi$ emitted during all three phases of shock expansion. The three lines in \Eq~(\ref{eq:factor0}) correspond to layers $\xi$ produced in different phases:
(I) $\Rx<\rem<R_1$ corresponds to $\xi_0<\xi<\xi_1=2\pi/\omega$,
(II) $R_1<\rem<\Rrad$ corresponds to $\xi_1<\xi<\xirad$, and 
(III) $\rem>\Rrad$ corresponds to $\xi>\xirad$. 
We also note that $\xirad$ is only slightly larger than $\xi_1$ ($\xirad\approx 1.03\,\xi_1$ in Model~W), and $\Rrad/R_1\sim 2$.

Examining now the result for $b/a$ (\Eq~\ref{eq:b_a1}),
\beq
\label{eq:b_a2}
   \frac{b}{a}(r,\xi)\approx \frac{\xiB^{1/2}}{18} S^2(r,\xi),
\eeq
with $S$ given by \Eq~(\ref{eq:factor0}), one can see that $b/a$ grows $\propto r^4$ at each $\xi$ as the FRB propagates at $r>\Rrad$. One can also see that at a given $r$ the profile of $b/a$ in $\xi$ has two peaks. The highest peak is near $\xi_0=3\pi/2\omega$ ($\rem\approx\Rx$), and the second peak is at $\xirad$, where the $\xi$-dependence of $S$ switches from $S\propto\rem^{5/4}$ to $S\propto \rem^{-2}$. 

Ratio $b/a$ exceeds 1/3 first in the leading part of the FRB, at $\xi\approx\xi_0$, where $S$ has the large factor $(1-\Rx^4/\rem^4)^{-1/4}$. This factor has its maximum value $\sim (c\sigx/\omega\Rx)^{1/4}$ (\Eq~\ref{eq:min}), about $30$ in Model~W. Once stochastic heating is activated in the leading part, this part is quickly absorbed. Thus, the FRB first becomes absorbed at $\xi<\xia$ (but not at $\xi>\xia$), where $\xia(r)$ is found from $b/a=1/3$ using the first line of \Eq~(\ref{eq:factor0}). This absorption front $\xia(r)$ is launched soon after the FRB radius $r$ crosses $\Rrad$. 

Then, at a larger $r$, FRB absorption becomes activated at $\xirad$, where $S$ (and hence $b/a$) has the second peak. The onset of absorption at $\xirad$ happens when the FRB reaches radius $r=\Rabs(\xirad)$ defined by
\beq 
\label{eq:b_a_rad}
  \frac{b}{a}(\xirad,r)=\frac{1}{3}.
\eeq
This absorption damages the main part of the FRB, erasing it at $\xi>\xib$, with $\xib(r)$ decreasing from $\xirad$ as the FRB expands to $r>\Rabs(\xirad)$.

The entire FRB is absorbed when the growing $\xia(r)$ and the decreasing $\xib(r)$ meet at $\xi_1=2\pi/\omega$. This occurs at $r=\Rabs(\xi_1)$ defined by 
\beq 
\label{eq:b_a_1}
  \frac{b}{a}(\xi_1,r)=\frac{1}{3},
\eeq
where $\xi_1=2\pi/\omega$ corresponds to $\rem=R_1$. 

The critical radius $\Rabs(\xirad)$, where main FRB absorption develops, can be found from \Eq~(\ref{eq:b_a_rad}) using $S$ evaluated at $\rem=\Rrad$ by extrapolating line (III) in \Eq~(\ref{eq:factor0}) to $\Rrad$. Then, \Eq~(\ref{eq:b_a_rad}) becomes
\beq
   \frac{\N(r) r^2}{\N(\Rrad)\Rrad^2}\approx \frac{\sqrt{6}}{\xiB^{1/4}}\approx \sqrt{60}.
\eeq
For instance, suppose the density parameter $\N$ satisfies $\N(r)\propto r^{-\alpha}$ with some slope $\alpha$. It gives
\beq
  \Rabs(\xirad)\approx \left(\frac{\sqrt{6}}{\xiB^{1/4}}\right)^{1/(2-\alpha)}\Rrad.
\eeq
Similarly, $\Rabs(\xi_1)$ (which is outside $\Rabs(\xirad)$) can be found from \Eq~(\ref{eq:b_a_1}) using $S$ evaluated at $\rem=R_1$ by extrapolating line (II) in \Eq~(\ref{eq:factor0}) to $R_1$.

Consider now Model~W that could explain FRBs from SGR~1935+2154. The model has $\N(\Rrad)=10^{37}$ and $\Rrad\approx 1.12\times 10^9$\,cm. Let us denote $\Rabs(\xirad)$ as $\Rabs$, omitting $\xirad$. If $\alpha=0$ (uniform $\N$), we find $\Rabs\approx 3\Rrad$, well inside the light cylinder of SGR~1935+2154, $\RLC\approx 1.55\times 10^{10}$\,cm, so the FRB does not escape. If $\alpha\sim 1$,  $\Rabs$ is pushed to the light cylinder, and the FRB  propagates through the closed magnetosphere without activating stochastic heating, avoiding strong damping.  We find that $\Rabs=\RLC$  if $\alpha\approx 1.22$, which corresponds to $\NLC\equiv \N(\RLC)\approx 4\times 10^{35}$. 

Note that our calculation of $b/a$ assumed $\c\ll 1$ ($\bD\approx -1$) which gives $\Ep\approx -\Bbg/2$. This approximation was used to evaluate $\gamma_0(\xi)$ on the plateau (\Eq~\ref{eq:energy}). At large $r\sim 10^{10}$\,cm, the condition $\c\ll 1$ becomes violated. This can be seen using $\c\approx a/2\gamma_0$, which gives 
\beq
  \c(r,\xi)\approx \frac{\aem\rem}{2\sigbg(r)\, c(\xi-\xiin)}.
\eeq
In particular, in Model~W, $\c(r,\xirad)$ exceeds unity where $\sigbg\lesssim 9\times 10^3$, which occurs when $r\N_{36}^{1/3}\gtrsim 8 \times 10^9$\,cm. Even when this happens, the value of $\gamma_0(r,\xirad)$ is weakly affected because $\gamma_0\propto (\xi-\xiin)$ grows over the extended interval $\xiin<\xi<\xirad$ while $\aem(\xi)$ is large enough to give $\c>1$ only near the end of this interval, at $\xi-\xiin\sim (2\nu)^{-1}$.

\subsection{FRB propagation in the wind at $r>\RLC$}

The wind is described by its power $\Lw$ (dominated by Poynting flux) and net particle flux $\dot\N$:
\beq
  \Lw\approx\frac{c\mu^2}{\RLC^4}, \qquad \dot\N\approx 4\pi c\,\frac{\NLC}{\RLC},
\eeq
where $\NLC\equiv \RLC^3\nbg(\RLC)$. The magnetic dipole moment $\mu$ and the wind power $\Lw$ are determined by the observed rotation period $P$ and its time derivative $\dot P$. SGR~1935+2154 has $\RLC=c P/2\pi\approx 1.55\times 10^{10}$\,cm and $\Lw\approx 2\times 10^{34}$\,erg\,s$^{-1}$. The density parameter $\NLC$ is not determined by observations and will be treated below as a free parameter. While its typical value in active magnetars is expected to be $\sim 10^{37}$ \citep{Beloborodov20}, $\NLC$ may vary. As shown above, FRBs emitted by monster shocks in SGR~1935+2154 escape the closed magnetosphere and enter the wind zone if $\NLC\lesssim 4\times 10^{35}$.

A key property of the wind is the high magnetization at $r=\RLC$:
\beq
  \sigLC \approx \frac{\mu^2}{4\pi\me c^2 \RLC^3 \NLC}\approx \frac{\Lw}{\dot\N\me c^2}.
\eeq
In particular, for SGR~1935+2154 $\sigLC\approx 10^3\N_{\rm LC,36}^{-1}$. The wind Lorentz factor may be approximated as growing linearly with radius until it saturates at $\sigLC^{1/3}$:
\beq
   \gbg\approx \frac{r}{\RLC}, \qquad \RLC<r<\sigLC^{1/3},
\eeq
so the wind quickly becomes ultrarelativistic, $\bbg\approx 1$. Its density $\nbg$ is determined by 
\beq
   \dot\N=4\pi r^2 c\bbg\nbg.
\eeq
Subscript ``bg'' highlights that the wind forms the background for the kHz magnetosonic wave and the FRB riding inside the wave. The large $\bbg$ means fast drift of (nearly transverse) magnetic field lines in the wind and implies a large electric field $\Ebg$. This field was negligible in the static closed magnetosphere, where we used $\Ebg\approx 0$ and $\Bbg\approx\mu/r^3$. The background electromagnetic field in the wind zone is 
\beq
  \Ebg=\bbg\Bbg, \qquad \Bbg\approx \frac{\mu}{\RLC^2 r}.
\eeq 

Consider now the propagating kHz magnetosonic wave crossing $\RLC$ and exiting into the wind zone. The wave profile has evolved inside the closed magnetosphere as described in \Sect~\ref{monster}. Its electric field has the plateau at $\xiin<\xi<\xish$, which continually accelerates ambient particles passing through the wave toward the shock. This acceleration effect continues in the wind zone. It is best viewed in the rest frame of the background wind $\KF'$, where 
\beq
   \Ebg'=0, \qquad \Bbg'=\frac{\Bbg}{\gbg}.
\eeq
In the frame $\KF'$, the picture of particle acceleration is similar to that in the closed magnetosphere (\Sect~\ref{monster}). The energy $(\gamma_0'-1)\me c^2$ gained by each plasma particle crossing the plateau is determined by the energy lost by the wave to keep the electromagnetic invariant $E'^2-B'^2$ below zero. The plateau with $E'^2\approx B'^2$ is sustained as long as the corresponding energy extracted from the wave is sufficient for particle acceleration to $\gamma_0'\gg 1$ (which corresponds to $\bD'=E'/B'\approx -1$). The kHz magnetosonic wave propagating $dr=cdt$ spends a well defined energy $d\E'$ (isotropic equivalent) to accelerate particles on the plateau \citepalias{Beloborodov23}:
\beq
  d\E' = -c(\xi'-\xiin') d(r^2\Ep'^2), \qquad \Ep'\approx -\frac{\Bbg'}{2},
\eeq
where $\xi'=\gbg(1+\bbg)\xi$; this transformation 
is found from $\omega\xi=inv$ and $\omega=\gbg(1+\bbg)\omega'$.

The energy $d\E'$ is received by $d\N$ particles that cross the plateau during time $dt=dr/c$:
\beq
  d\N=\dot\N(1-\bbg)\frac{dr}{c},
\eeq
and one can find $\gamma_0'$ from 
\beq
  (\gamma_0'-1)\me c^2d\N=d\E',
\eeq
which gives
\beq
\label{eq:g0_wind1}
  \gamma_0' \approx 1+\frac{\sigLC(1+\bbg) c(\xi-\xiin) }{\RLC}.
\eeq
Here, we used $d(r^2\Bbg'^2)/dr\approx -2r\Bbg'^2$ (which follows from $\Bbg'^2=\Bbg^2/\gbg^2\propto r^{-4}$ at $1<r/\RLC<\sigLC^{1/3}$) and then replaced $-2$ with $-4/(1+\bbg)$ to smoothly match (at $r=\RLC$) with $d(r^2\Bbg^2)/dr=-4 r\Bbg^2$ that is found at $r<\RLC$. Then, at $\bbg=0$ ($r=\RLC$), \Eq~(\ref{eq:g0_wind1}) matches the expression for $\gamma_0'=\gamma_0$ found in the closed magnetosphere (\Sect~\ref{monster}). 

The plateau accelerator can also be described by the quantity $\c_0'\equiv\gamma_0'(1+\beta_0')$, and we find 
\beq
\label{eq:kappa0_wind}
  \c_0'(\xi,r)\approx \left[ 1+\frac{2\sigLC(1+\bbg) c(\xi-\xiin)}{\RLC}\right]^{-1}.
\eeq
This expression holds in the zone $1<r/\RLC<\sigLC^{1/3}$ and gives $\c_0'$ approximately constant with $r$, as $1+\bbg$ saturates at 2. The plateau accelerator is efficient when $\gamma_0'\gg 1$ or, equivalently, when $\c_0'\ll 1$. The plateau extends to $\xi-\xiin\sim (2\nu)^{-1}$, nearly reaching $\xirad$ (see the last paragraph of section~\ref{stoch}), so one can use \Eq~(\ref{eq:kappa0_wind}) to estimate $\c_0'$ at $\xirad$. This gives 
\beq 
\label{eq:plateau_wind}
   \c_0'\approx \frac{\RLC\,\nu}{c\sigLC(1+\bbg)}\ll 1 \;\; {\rm when} \;\,  \sigLC\gg 200 \, \nu_3 R_{\rm LC,10}. 
\eeq
In particular, in SGR~1935+2154, one finds $\c_0'\ll 1$ if $\sigLC\gg 300\,\nu_3$, which corresponds to $\NLC\ll 3\times 10^{36}\nu_3^{-1}$. 

The corresponding parameter $\c_0\equiv\gamma_0(1+\beta_0)$ in the lab frame is given by the Lorentz transformation
\beq
   \c_0=\cbg\c_0',
\eeq
where $\cbg\equiv \gbg(1+\bbg)$. 

In the region occupied by the FRB, $\xi_0<\xi<\xish$, the MHD fluid drifts with speed $\bD$ that differs from $\beta_0$, because the FRB with strength parameter $a(\xi,r)\gg 1$ forces strong oscillations of the plasma particles in the local fluid frame $\tKF$ (\Sect~\ref{deceleration}). As long as the plasma responds in the regular oscillation regime, with a steady pattern of the fluid flow in the precursor, the parameter $\c\equiv\gD(1+\bD)$ satisfies $\c=\sqrt{1+a^2}\c_0$, and so
\beq
\label{eq:kappa_wind}
  \c=\sqrt{1+a^2}\,\cbg\c_0'. 
\eeq
Note that $\c\approx const$, since $\c_0'\approx const$, $\sqrt{1+a^2}\approx a\propto r^{-1}$, and $\cbg\propto r$ at $\RLC\ll r<\sigLC^{1/3}$. For example, our FRB model for SGR\,1935+2154 (Model~W) predicts $a(r)\approx 20\RLC/r$ at the peak of the burst (at $\xi\approx\xirad$), which gives $\c\approx 20 \c_0'\approx 8 \N_{\rm LC,36}$. 

The plasma responds to the FRB in the regular oscillation regime if $b/a>1/3$, where $b\equiv \tomB/\tompre$ and $\tomB=e\tB/\me c$. To evaluate $b$, first note that in a short magnetosonic wave ($c/\omega\ll r$), magnetic flux freezing condition gives $(1-\bD)B=(1-\bbg)\Bbg$, which implies
\beq
  \tB = \c(1-\bbg)\Bbg.
\eeq
Then, using $\tompre=\ompre/\c$, we find
\beq
\label{eq:b_wind}
   b\equiv\frac{\tomB}{\tompre} =\c^2(1-\bbg)\frac{\omB}{\ompre},    \qquad   \omB\equiv\frac{e\Bbg}{\me c}.
\eeq

One can now see that the calculation of $b/a$ in the wind zone repeats what was done in \Sect~\ref{stochastic} at $r<\RLC$ with $\Bbg$ replaced by $(1-\bbg)\Bbg$ and $\c_0$ replaced by $\c_0'$ found in the wind rest frame $\KF'$. \Eq~(\ref{eq:b_wind}) yields
\beq
\label{eq:b_a1_}
   \frac{b}{a}(r,\xi)= \c^2(1-\bbg)\frac{\Bbg}{\Erms} = \left(\frac{b}{a}\right)_{\rm em} \! S^2,
\eeq
where $(b/a)_{\rm em}\approx \xiB^{1/2}/18$ (\Eq~\ref{eq:b_a_em}) and
\beq
   S^2(r,\xi) \equiv  \frac{\c^2}{\cem^2} \frac{\aem}{a} \frac{(1-\bbg)\Bbg}{\Bbg^{\rm em}}.
\eeq
Using $\aem/a=r/\rem$, $\Bbg^{\rm em}=\mu/\rem^3$, $\Bbg=\mu/\RLC^2 r$, $\cem\approx \aem/2\gem$, and substituting $\c^2$ from \Eq~(\ref{eq:kappa_wind}), we obtain
\beq
   S\approx \frac{\rem^2\gem}{c(\xi-\xiin)\sigLC(1+\bbg)^{1/2}r}.
\eeq
We conclude that $S$ decreases in the wind zone as $r^{-1}$, and so $b/a$ decreases as $r^{-2}$. At $r<\RLC$, we found $S\propto r^{2}$ (\Eq~\ref{eq:factor0}) and $b/a\propto r^{4}$. Hence, $b/a$ is maximum near the boundary of the closed magnetosphere $r\approx\RLC$, and the condition $\Rabs>\RLC$ discussed in section~\ref{stoch} is sufficient for the FRB escape.

%##############################################################

\bibliography{ms}

\end{document}